\documentclass[letterpaper,10pt,namedreferences,nointegrals,linenumbers]{solarphysics}
\usepackage[hyperref,optionalrh,solaromanenum]{spr-sola-addons} 
\usepackage{graphicx}                    
\usepackage{amsmath}
\usepackage{amssymb}                    
\usepackage[dvipsnames]{xcolor}
\usepackage{bm}
\usepackage{enumitem}

\renewcommand{\vec}[1]{ {\bm #1} }
\newcommand{\uvec}[1]{ \hat{\bm #1} }

\defcitealias{fan2003}{FG03}

\begin{document}

\begin{article}

\begin{opening}

\title{Non-Neutralized Electric Current of Active Regions Explained as a Projection Effect}


\author[addressref={uhifa},corref,email={xudongs@hawaii.edu}]{\inits{X.}\fnm{Xudong}~\lnm{Sun}\orcid{0000-0003-4043-616X}}
\author[addressref={lmsal}]{\inits{M.~C.~M.}\fnm{Mark~C.~M.}~\lnm{Cheung}\orcid{0000-0003-2110-9753}}

\runningauthor{Sun \& Cheung}
\runningtitle{Current Neutralization}

\address[id=uhifa]{Institute for Astronomy, University of Hawai`i at M\={a}noa, Pukalani, HI 96768, USA}
\address[id=lmsal]{Lockheed Martin Solar and Astrophysics Laboratory, Palo Alto, CA 94304, USA}


\begin{abstract}

Active regions (ARs) often possess an observed net electric current in a single magnetic polarity. We show that such ``non-neutralized'' currents can arise from a geometric projection effect when a twisted flux tube obliquely intersects the photosphere. To this end, we emulate surface maps of an emerging AR by sampling horizontal slices of a semi-torus flux tube at various heights. Although the tube has no net toroidal current, its \textit{poloidal} current, when projected along the \textit{vertical} direction, amounts to a significant non-neutralized component on the surface. If the tube emerges only partially as in realistic settings, the non-neutralized current will 1) develop as double ribbons near the sheared polarity inversion line, (2) positively correlate with the twist, and 3) reach its maximum before the magnetic flux. The projection effect may be important to the photospheric current distribution, in particular during the early stages of flux emergence.

\end{abstract}



\end{opening}

%

\section{Introduction}
\label{sec:intro}


\subsection{Definitions of Current Neutralization}
\label{subsec:definition}

Magnetic field in the solar interior is believed to exist in a ``fibril'' state: isolated flux tubes generated from dynamo actions are embedded in a relatively field-free background \citep{fan2009lrsp}. Strong field strength and twist keep the tubes coherent against vigorous convection \citep{schuessler1979}. This places an interesting constraint on the net electric current $I$. Using Amp\`{e}re's law and Stokes theorem, we have
\begin{equation}
\label{eq:stokes}
\begin{split}
\vec{j} &= \mathbf{\nabla} \times \vec{B}, \\
I	& = \int_C \vec{j}\cdot\uvec{n}\,\mathrm{d} A = \oint_{\partial C} \vec{B}\cdot\mathrm{d}\vec{l}, \\
\end{split}
\end{equation}
where $\vec{j}$ is the electric current density, $\vec{B}$ is the magnetic field, $A$ is an area on the tube cross section, $\uvec{n}$ is its unit normal vector, $C$ is any surface with area $A=\int_C dA$, $\partial C$ is the perimeter of $C$, and $\vec{l}$ is its tangent vector. For simplicity, we assume unity magnetic permeability. Suppose a magnetic flux tube is compact in space. More specifically, there is a closed loop $\partial C$ such that $B=0$ all along the loop. This shows that a compact flux tube has $I=0$, i.e. it is current-neutralized. 

The argument implies \textit{neutralized} electric currents in the axial direction. For a tube with coherent twist, $\vec{j}$ near the tube center will point in one direction when projected onto the axis. Conversely, $\vec{j}$ in the tube periphery must point in the opposite direction, forming a sheath or a skin layer. The former is termed the ``direct current'' (DC), and the latter the ``return current'' \citep[RC; e.g.,][]{melrose1995}. The flux tube is ``current neutralized'' when DC and RC add to zero.

Solar active regions (ARs), formed through emergence of subsurface flux tubes, are known to harbor electric currents in the corona. In many cases, their sheared or sigmoidal loops are incompatible with a current-free, potential field morphology. Moreover, they must contain sufficient free energy to power flares and coronal mass ejections (CMEs). This picture is quite different from the convection zone. Here, the magnetic field is volume-filling and dominates the plasma dynamics. 

In the thin layer of the photosphere, the plasma density and pressure decrease rapidly with height; the magnetic field begins to transition from a fibril to a volume-filling state. Maps of the photospheric $\vec{B}$, magnetograms, are routinely inferred from spectropolarimetric observations. Assuming a local Cartesian geometry and the photosphere as a thin boundary at $z=0$, the \textit{vertical} current density $j_z$ and the net current $I_z$ can be estimated from the horizontal field component $\vec{B}_h$, similar to Equation~(\ref{eq:stokes}):
\begin{equation}
\label{eq:stokesph}
\begin{split}
j_z &= \mathbf{\nabla} \times \vec{B}_h, \\
I_z	& = \int_C j_z \, \mathrm{d}A = \oint_{\partial C} \vec{B}_h\cdot\mathrm{d}\vec{l}, \\
\end{split}
\end{equation}
where $C$ is now a finite area lying entirely on $z=0$.

For an isolated, spatially compact sunspot, there exists a closed loop $\partial C$ lying on $z=0$ that surrounds the sunspot such that $\vec{B}(z=0)=0$. This implies $I_z = 0$. That is, all currents that enter the photosphere must return to the convection zone, i.e., they are ``balanced''. In practice, however, this is not how net currents are measured for observed ARs. First of all, sunspots are rarely completely isolated. Furthermore, eruptive ARs tend to harbor sunspots of opposite polarities in close proximity. To measure the net current, standard practice is to take $C$ as the contiguous area of a magnetic polarity partially bounded by the polarity inversion line (PIL; i.e. $B_z = 0$). Figure~1(a) of \citet{torok2014} shows a good example. DC and RC are expected to reside in the core region and the periphery, respectively.


\subsection{Non-Neutralized Currents in ARs}
\label{subsec:literature}

Ground-based vector magnetograms provided early evidence for non-neutralized currents in the AR photosphere. \citet{leka1996} followed the evolution of an emerging AR over several days. A significant net current appears in each magnetic polarity, whose increase cannot be explained by the shearing flows in the photosphere alone and must be carried by the emerging flux. \citet{wheatland2000} analyzed a sample of 21 ARs. In most cases, the net current for the entire AR is consistent with zero. Within each magnetic polarity, however, it is significantly different from zero.

Using high-resolution vector magnetograms from the Spectropolarimeter (SP) on board the \textit{Hinode} satellite, \citet{ravindra2011} and \citet{georgoulis2012} found that the non-neutralized currents are always accompanied by well-formed, sheared PILs. Meanwhile, \citet{venkatakrishnan2009} and \citet{gosain2014} found isolated sunspots without a sheared PIL to be well neutralized. This is a natural consequence of Equation~(\ref{eq:stokesph}): the sheared component of $\vec{B}_h$ along the PIL should contribute significantly to the line integral around a single magnetic polarity.

The current distribution has perceived important implications for solar eruptions. Many CME models employ DC-dominated flux ropes as the driver \citep[e.g.,][]{titov1999}. The inclusion of RC will reduce the outward magnetic ``hoop force'' that drives the eruption \citep{forbes2010}. Observationally, non-neutralized currents indeed correlate with eruptive activities. \citet{liuy2017}, \citet{vemareddy2019}, and \citet{avallone2020} compared ARs with and without major eruptions using data from the Helioseismic and Magnetic Imager (HMI) aboard the \textit{Solar Dynamics Observatory} (\textit{SDO}). The quiescent ARs are mostly current-neutralized, while the active ones are not. \citet{falconer2001} and \citet{kontogiannis2019} demonstrated the net current as a useful index for space weather predictions. On a related note, electric currents can also cause an apparent line-of-sight flux imbalance because of the directionality of the magnetic field they produce \citep{gary1995}.

The origin of the non-neutralized current in ARs is not entirely clear. If flux emergence is responsible, it must explain how it can arise from a current-neutralized subsurface flux tube. \citet{longcope2000} proposed an analytical model where only a fraction of the current passes into the corona. The rest, including most of the weaker RC, is shunted to the surface layers and hidden from the observer. \citet{torok2014} demonstrated a similar effect using a magnetohydrodynamic (MHD) simulation from \citet{leake2013}. As a subsurface, neutralized flux tube emerges, non-neutralized current develops simultaneously with the flux emergence and PIL shear. The degree of non-neutralization becomes more pronounced in higher layers as RC remains trapped in lower ones.

Shear and vortex surface flows are often used to induce coronal electric currents in MHD models. \citet{torok2003} and \citet{dalmasse2015} showed that different photospheric line-tied motions can generate different degrees of current neutralization. Only motions that directly shear the PIL will generate a non-neutralized component. Simulations by \citet{fan2001} and \citet{manchester2004} showed that the shear flows on the PIL are driven by the magnetic tension force during flux emergence, which comes from the twisted tube itself as it expands in the corona.


\subsection{Outline}
\label{subsec:outline}

Here we explore an alternative origin of the observed, non-neutralized electric current in emerging ARs. Using an analytical flux tube model, we show that a neutralized subsurface flux tube can lead to a significant non-neutralized vertical current on the surface due to a geometric projection effect.

We described two versions current neutralization in Section~\ref{subsec:definition}: one in the context of the subsurface flux tube itself, and the other in the context of the photospheric observation. As we shall see, they refer to different components of the current system when the tube axis obliquely intersects the surface. Such a distinction is seldom discussed in the literature, but is central to this work. 

Below, we describe the flux tube model in Section~\ref{sec:model}, present our results in Section~\ref{sec:result}, and discuss the implications in Section~\ref{sec:discussion}. Additional details of the flux tube model are available in the appendix.


\begin{figure}[t!]
\centerline{\includegraphics[width=0.94\textwidth]{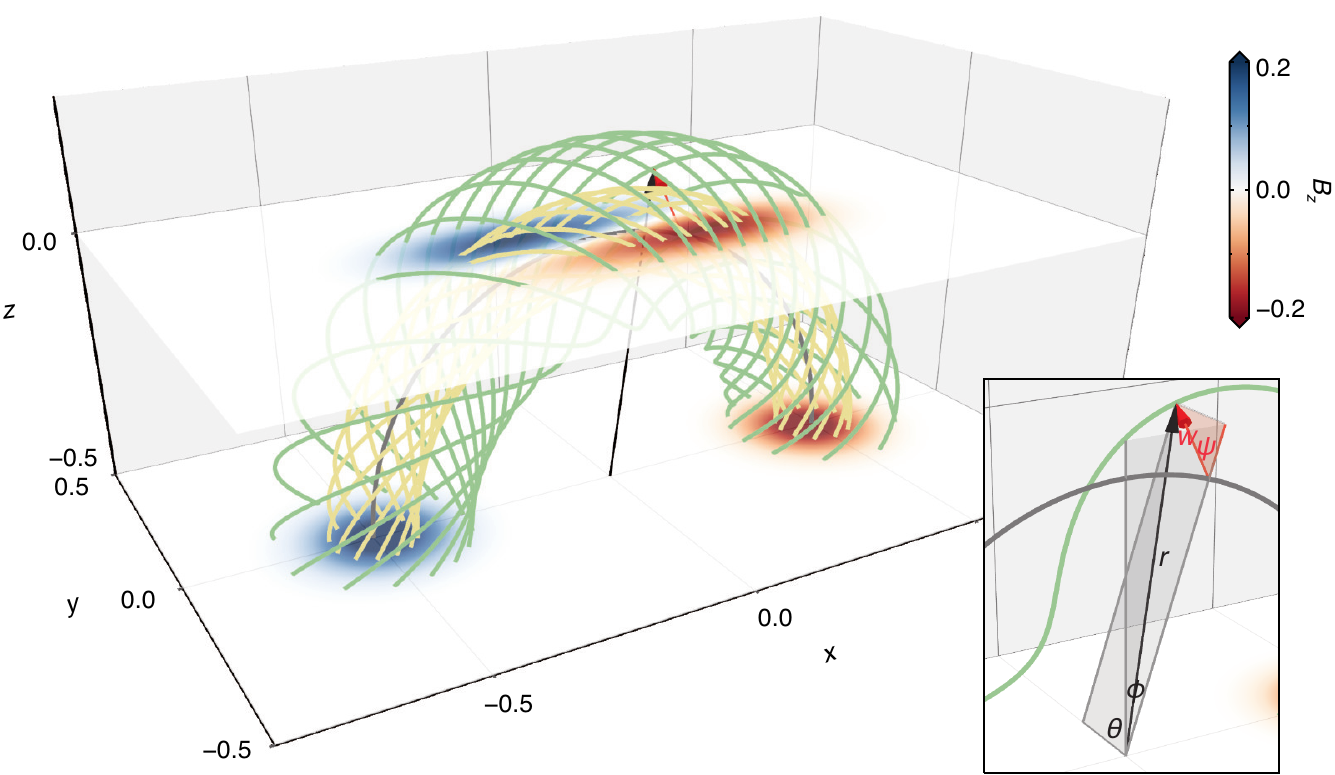}}
\caption{Magnetic field of a partially emerged \citetalias{fan2003} flux tube ($R=0.5$, $a=0.12$, $q=0.8$, $h=-0.5$). The tube axis (dark gray curve) just starts to touch the photosphere at $z=0$. Maps on two horizontal slices at $z=-0.5$ and $0$ show $B_z$ distribution. Selective field lines are traced from a distance of $w=0.08$ (yellow) and $0.18$ (green) from the tube axis in the $x=0$ plane. They are located within the DC core and RC sheath, respectively. The black and red arrows show the radial vector for an arbitrary point in the spherical and torus coordinate system, respectively (see appendix). The inset illustrates the spherical coordinate $(r,\theta,\phi)$, and the torus coordinate $(w,\psi,\phi)$. The poloidal angle $\psi$ here is negative.}
\vspace{2mm}
\label{f:mfr}
\end{figure}



\section{Flux Tube Model}
\label{sec:model}


\subsection{Model Setup}
\label{subsec:setup_fg03}

We consider a twisted, semi-torus flux tube model from \citet[][hereafter FG03]{fan2003}. In a global Cartesian coordinate with unit vectors $(\uvec{x}, \uvec{y}, \uvec{z})$, the subsurface tube is centered at $\vec{r}_0=(0,0,h)$, where $h<0$ (Figure~\ref{f:mfr}). 

The semi-torus tube axis has a radius of $R$; it is axisymmetric with a polar axis parallel to $\uvec{y}$. In a local spherical coordinate centered at $\vec{r}_0$ with unit vector $(\uvec{r}, \uvec{\theta}, \uvec{\phi})$, the magnetic field $\vec{B}$ inside the tube is defined as
\begin{equation}
\label{eq:bsph}
\vec{B} = \mathbf{\nabla} \times \left[ \dfrac{A(r, \theta)}{r \sin{\theta}} \, \uvec{\phi} \right] + B_\phi(r, \theta) \, \uvec{\phi},
\end{equation}
where
\begin{equation}
\label{eq:bsph0}
\begin{split}
A(r,\theta) & = \dfrac{q a^2 B_t}{2}  \exp\left(-w^2/a^2\right), \\
B_\phi(r,\theta) & = \dfrac{a B_t}{r \sin{\theta}} \exp\left(-w^2/a^2\right). \\
\end{split}
\end{equation}
Here, $r$ is the radial distance from the origin; $\theta$ is the polar angle from the polar axis; $\phi$ is the azimuth angle where 0 is along $+z$ and increases clockwise toward $+x$ ($-\pi/2 \le \theta, \phi \le \pi/2$). Moreover, $w$ is the distance from the tube axis. $B_t$, $q$, and $a$ are constants: $B_t$ is the characteristic field strength; $q$ is a measure of the twist; $a$ is the characteristic length scale for the tube radius. Field lines near the tube axis wind about the axis at a rate of $q/a$ per unit length. We truncate $\vec{B}$ to 0 outside the surface $w=3a$. The tube is thus isolated with a circular cross section and a radius of $3a$. Detailed expressions of $\vec{B}$ and $\vec{j}$ in various coordinate systems are available in the appendix.

In this work, we consider the following default parameters: $B_t=1.0$, $R=0.5$, and $a=0.12$. Additionally, we limit ourselves to the cases where $q>0$, such that the flux tube has a right-handed twist.

To emulate flux emergence, we displace the flux tube upward by increasing $h$ toward zero, and evaluate $\vec{B}$ and $\vec{j}$ on the horizontal slices $z=0$. This is equivalent to sampling the variables at different heights if the flux tube is fixed in space. The process is demarcated by several key values of $h$:
\begin{itemize}[noitemsep,topsep=4pt,parsep=2pt,leftmargin=16pt]
\item The emergence starts at $h=-(R+3a)=-0.86$.
\item The axis of the tube reaches the surface at $h=-R=-0.5$.
\item The crest of the tube detaches from the surface at $h=-(R-3a)=-0.14$.
\item The emergence ends at $h=0$.
\end{itemize}

In \citetalias{fan2003}, the tube field at $z=0$ was used to prescribe a time-dependent boundary condition that drives the coronal dynamics. Here, we focus solely on the layer $z=0$ as a proxy of the photospheric observations, ignoring any coronal evolution or its feedback. This approach has been used to help interpret the emerging non-potential field structure observed in the photosphere with success \citep{lites1995,gibson2004,luoni2011}.


\subsection{Poloidal-Toroidal Decomposition}
\label{subsec:ptd_fg03}

The axial direction of a torus is known as the \textit{toroidal} direction, which is just $\uvec{\phi}$ in our spherical coordinate. The non-axial direction, on the other hand, includes both $\uvec{r}$ and $\uvec{\theta}$. We may define a torus coordinate where a non-axial vector can be decomposed into radial and \textit{poloidal} components (see Figure~\ref{f:mfr} and appendix). In a constant $\phi$ plane, the former directs away from the tube axis, while the latter follows a circle around the tube axis.

We show in the appendix that the non-axial $\vec{B}$ and $\vec{j}$ in the \citetalias{fan2003} flux tube are purely poloidal. Using the subscript P (T) to denote the poloidal (toroidal) components, we have in spherical coordinate
\begin{equation}
\label{eq:btp}
\begin{split}
\vec{B}_\mathrm{P} & = B_r \uvec{r} +  B_\theta \uvec{\theta}, \\
\vec{B}_\mathrm{T} & = B_\phi \uvec{\phi}, \\
\end{split}
\end{equation}
\begin{equation}
\label{eq:jtp}
\begin{split}
\vec{j}_\mathrm{P} &= j_r \uvec{r} + j_\theta \uvec{\theta}, \\
\vec{j}_\mathrm{T} &= j_\phi \uvec{\phi}. \\
\end{split}
\end{equation}
Their respective contributions to the surface observation vary at different stages of the flux emergence, when the tube axis intersects the surface at different angles. In particular, the projected values in the vertical direction can be evaluated by taking the dot product with $\uvec{z}$, e.g., $\vec{B}_\mathrm{P} \cdot \uvec{z}$.


\subsection{Two Versions of Current Neutralization}
\label{subsec:dcrc_fg03}

Considering the flux tube by itself, we define DC and RC as components of the net toroidal current. For $q>0$, DC (RC) refers to the net current for $j_\phi>0$ ($j_\phi<0$). We denote the total DC as $I_\mathrm{T}^\mathrm{D}$, the total RC as $I_\mathrm{T}^\mathrm{R}$, and the net toroidal current $I_\mathrm{T}$. We can evaluate them by integrating $j_\phi$ on a cross section $\phi=0$:
\begin{equation}
\label{eq:itdr}
\begin{split}
I^\mathrm{D}_\mathrm{T} &= \int_{\phi=0} \left. j_\phi \right|_{j_\phi>0} \, \mathrm{d}A, \\
I^\mathrm{R}_\mathrm{T} &= \int_{\phi=0} \left. j_\phi \right|_{j_\phi<0} \, \mathrm{d}A, \\
I_\mathrm{T} &= I_\mathrm{T}^\mathrm{D}+I_\mathrm{T}^\mathrm{R}. \\
\end{split}
\end{equation}

For a photospheric vector magnetogram, we denote the total DC in a single magnetic polarity as $I^\mathrm{D}_z$, the total RC as $I^\mathrm{R}_z$, and the net vertical current $I_z$. For $q>0$, we can evaluate them by integrating $j_z$ on $z=0$ where $B_z>0$:
\begin{equation}
\label{eq:izdr}
\begin{split}
I^\mathrm{D}_z &= \int_{z=0;B_z>0} \left. j_z \right|_{j_z>0} \, \mathrm{d}A, \\
I^\mathrm{R}_z &= \int_{z=0;B_z>0} \left. j_z \right|_{j_z<0} \, \mathrm{d}A, \\
I_z &= I^\mathrm{D}_z+I^\mathrm{R}_z. \\
\end{split}
\end{equation}
We can separately evaluate the projected poloidal and toroidal contribution by integrating $\vec{j}_\mathrm{P}\cdot\uvec{z}$ and $\vec{j}_\mathrm{T}\cdot\uvec{z}$ instead of $j_z$.

The degree of current neutralization are often quantified by the relative magnitude of DC versus RC. In this study, we use the index
\begin{equation}
\label{eq:rt}
R_\mathrm{T} = \left| \dfrac{I^\mathrm{D}_\mathrm{T}}{I^\mathrm{R}_\mathrm{T}} \right|
\end{equation}
for the toroidal current and
\begin{equation}
\label{eq:rz}
R_z = \left| \dfrac{I^\mathrm{D}_z}{I^\mathrm{R}_z} \right|
\end{equation}
for the observed vertical current \citep[e.g.,][]{liuy2017,avallone2020}. We will show that a geometric projection effect is capable of producing large $I_z$ and $R_z$ from a flux tube with small $I_\mathrm{T}$ and unity $R_\mathrm{T}$.


\section{Result}
\label{sec:result} 


\subsection{Flux Tube Cross Section}
\label{subsec:res_ft}

\begin{figure}[t!]
\centerline{\includegraphics[width=1.0\textwidth]{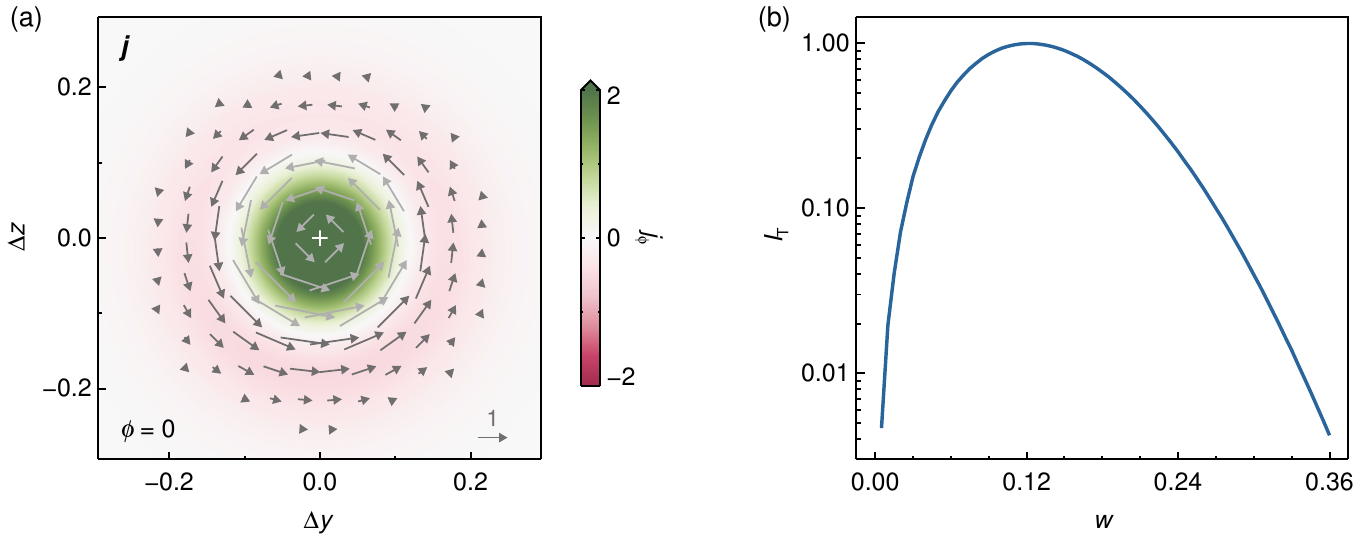}}
\caption{Electric current on a tube cross section ($q=0.8, \phi=0$). (a) Distribution of $\vec{j}$ near the tube axis. The background color shows the toroidal current density $j_\mathrm{T}$; the arrows show poloidal current density vectors $\vec{j}_\mathrm{P}$. The tube axis is marked with a cross. The coordinate $(\Delta y, \Delta z)$ indicates the offset from the tube axis in $y$ and $z$ direction. (b) Net toroidal current $I_\mathrm{T}$ within a tube radius $w$, normalized by $I^\mathrm{D}_\mathrm{T}$ of the entire tube.}
\label{f:itor}
\end{figure}

For our fiducial case with $q=0.8$, the toroidal current density on the tube cross section features a compact DC core and a diffuse RC sheath (Figure~\ref{f:itor}(a)). Though not strictly axisymmetric because of the bending of the flux tube axis, the sign change occurs near the radius $w \approx a = 0.12$ (Equation~(\ref{eq:jsph3})). The net toroidal current integrated within a circle of radius $w$ peaks at $w = 0.12$ due to the contribution from the DC core. It becomes quite neutralized once the RC sheath is included (Figure~\ref{f:itor}(b)). For the entire tube $w=3a=0.36$, there is $I_\mathrm{T}=0.004 I_\mathrm{T}^\mathrm{D}$ and $R_\mathrm{T}=1.004$. Strictly speaking, an additional RC skin layer with a net current of $-0.004 I_{\mathrm{T}}^{\mathrm{D}}$ is required. We ignore this layer as it does not affect any of our conclusions. Because $I_\mathrm{T}^\mathrm{D}$ and $I_\mathrm{T}^\mathrm{R}$ both linearly scale with $q$, $R_\mathrm{T}$ is independent of the twist.

In contrast, the poloidal current density has a single sign, rotating counter-clockwise around the axis in the same direction as the poloidal field (Equation~(\ref{eq:jtoi3}); Figure~\ref{f:itor}(a)). Its magnitude is independent of $q$ by construct, and is comparable to the toroidal current density for $q=0.8$.


\begin{figure}[t!]
\centerline{\includegraphics[width=1.0\textwidth]{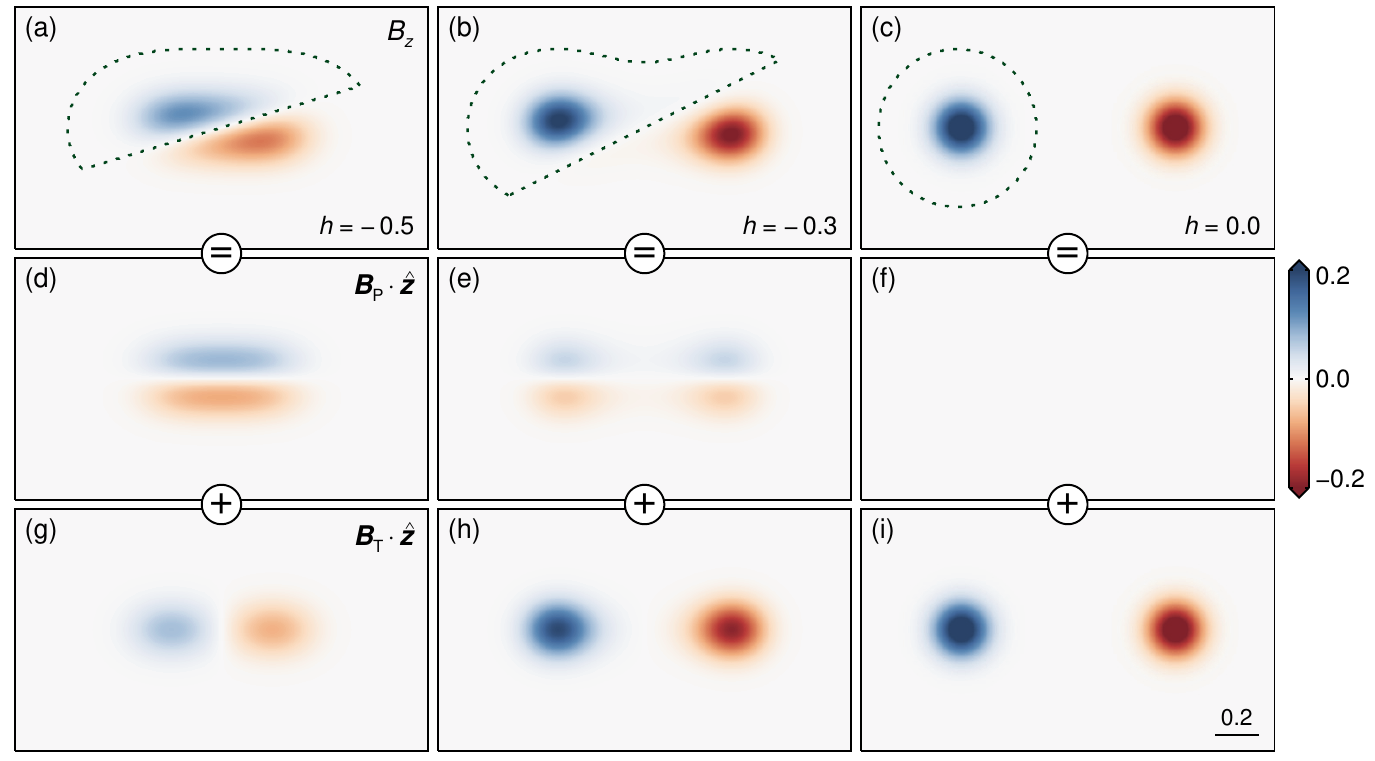}}
\caption{Surface magnetic maps ($q=0.8$, $z=0$). (a)--(c) Vertical magnetic field $B_z$ when the tube is centered at $h=-0.5$,~$-0.3$, and~$0$, respectively. Dotted curves indicate regions where $B_z>0$. (d)--(f) Contribution from the poloidal field $\vec{B}_\mathrm{P}\cdot\uvec{z}$. (g)-(i) Contribution from the toroidal field $\vec{B}_\mathrm{T}\cdot\uvec{z}$. The top row is the sum of the middle and bottom rows. For corresponding current density maps, see Figure~\ref{f:mapsj}.}
\label{f:mapsb}
\end{figure}


\begin{figure}[t!]
\centerline{\includegraphics[width=1.0\textwidth]{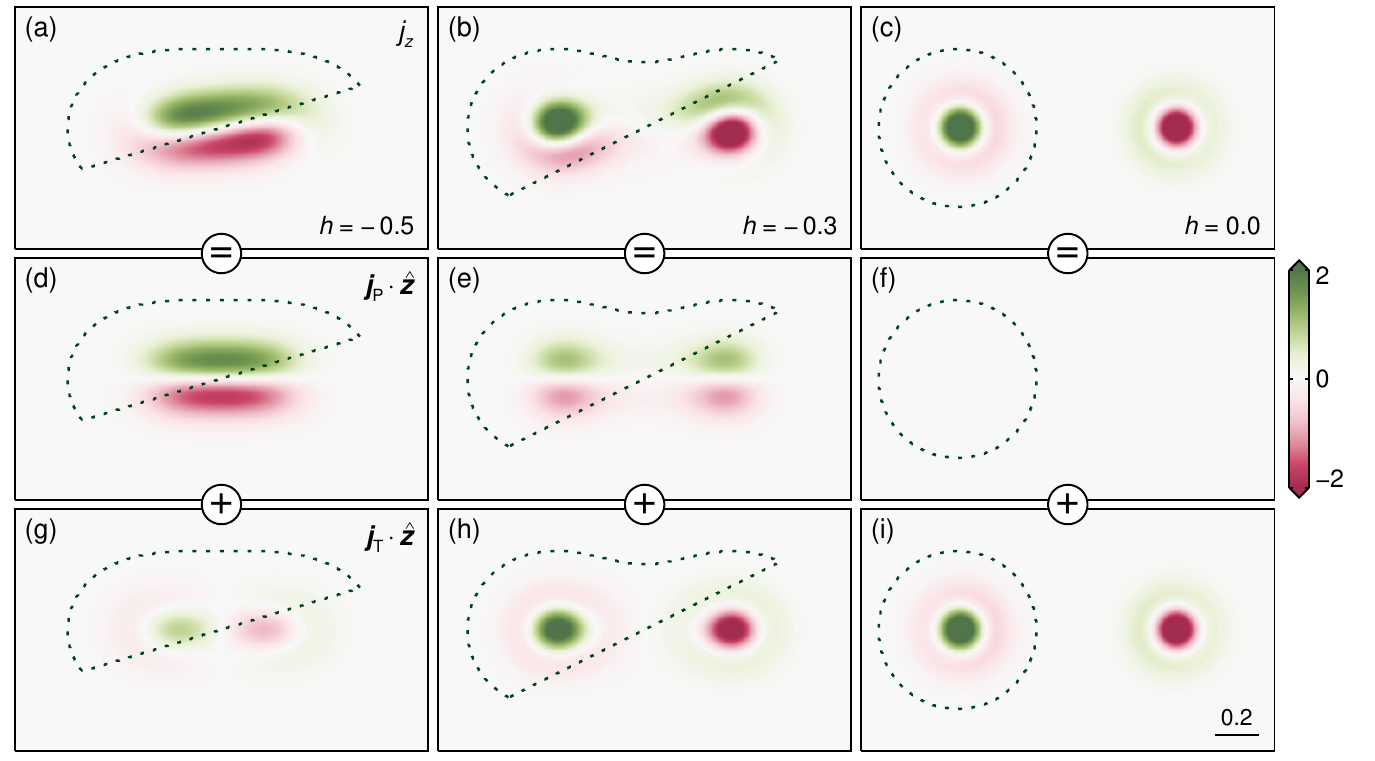}}
\caption{Surface current density maps ($q=0.8$, $z=0$). (a)--(c) Vertical electric current density $j_z$ when the tube is centered at $h=-0.5$,~$-0.3$, and~$0$, respectively. (d)--(f) Contribution from the poloidal current $\vec{j}_\mathrm{P}\cdot\uvec{z}$. (g)-(i) Contribution from the toroidal current $\vec{j}_\mathrm{T}\cdot\uvec{z}$. Dotted curves indicate regions where $B_z>0$. The top row is the sum of the middle and bottom rows.}
\label{f:mapsj}
\end{figure}


\subsection{Synthetic Surface Observations}
\label{subsec:res_surf}

The vertical magnetic field $B_z$ shows two elongated polarities when the crest of the tube intersects the surface (Figure~\ref{f:mapsb}(a)). These are known observationally as ``magnetic tongues'', which are a manifestation of the magnetic twist \citep{lopezfuentes2000}. As $h$ increases, the tongues retract and transition into two circles, which correspond to the two legs of the flux tube (Figure~\ref{f:mapsb}(b)-(c)). The poloidal and toroidal contributions to the vertical field are more important in the early and late stages, respectively (Figure~\ref{f:mapsb}(d)-(i)). 

The vertical current density $j_z$ shows two elongated, ribbon-shaped polarities early on (Figure~\ref{f:mapsj}(a)). They mainly arise from the projected poloidal component (Figure~\ref{f:mapsj}(d)). For $q=0.8$, $j_z$ is mostly positive in the positive magnetic polarity, so DC dominates. As the tube continues to rise, the RC sheath becomes visible on the surface, and the $j_z$ distribution starts to resemble $j_\phi$ on the tube cross section (Figure~\ref{f:mapsj}(g)-(i)).


\begin{figure}[t!]
\centerline{\includegraphics[width=0.99\textwidth]{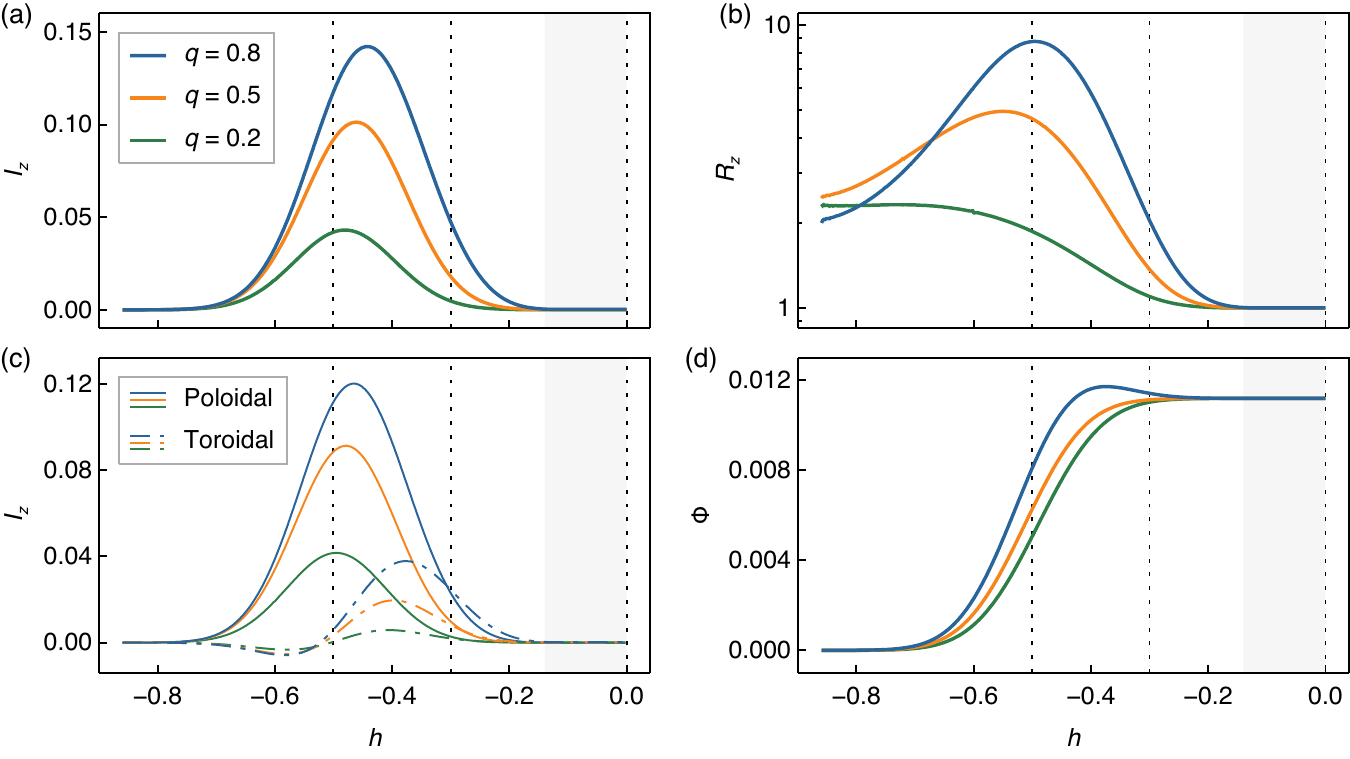}}
\caption{Integrated surface observations for different $q$. (a) Net current $I_z$. (b) Ratio between DC and RC $R_z$. (c) Projected poloidal (thin solid) and toroidal (dash-dotted) components in $I_z$. (d) Surface magnetic flux $\Phi$ ($B_z > 0$ only). Gray bands indicate $h\ge-(R-3a)$. Dotted lines show the three heights in Figure~\ref{f:mapsj}.}
\label{f:profile}
\end{figure}


\begin{figure}[t!]
\centerline{\includegraphics[width=1.0\textwidth]{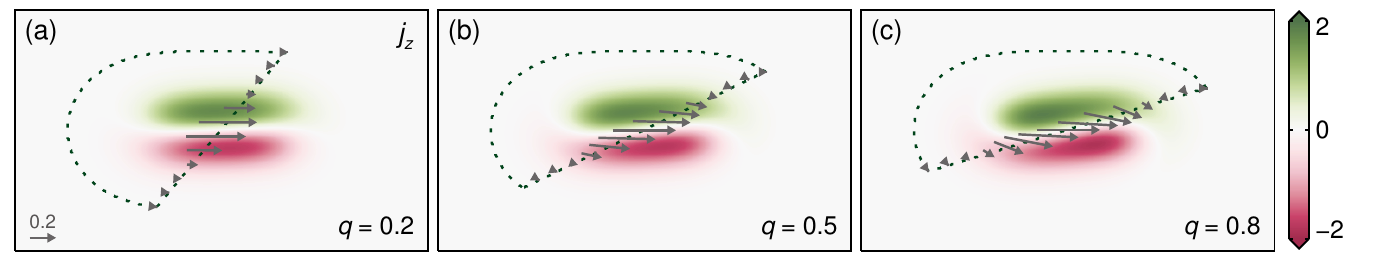}}
\caption{Surface distribution of $j_z$ for $h=-0.5$, when the apex of the flux rope axis is at the photospheric level. (a)-(c) For $q=0.2$, $0.5$, and $0.8$, respectively. Dotted curves indicate regions where $B_z>0$. Arrows show $\vec{B}_h$ on the PIL.}
\label{f:jz}
\end{figure}


For $q=0.8$, the surface vertical current is highly non-neutralized when the axis just starts to emerge ($h=-R$). $I_z$ and $R_z$ peak at $h=-0.44$ and $-0.50$, with a maximum of $I_z=2.6I_\mathrm{T}^\mathrm{D}$ and $R_z=8.8$, respectively (Figure~\ref{f:profile}(a)-(b)). The dominance of DC is clearly owing to the poloidal contribution (Figure~\ref{f:profile}(c)). At $h=-0.5$, the contribution of the poloidal component in $I_z$ is about $18$ times that of the toroidal counterpart. 

To evaluate the general properties of the model, we consider two additional cases for $q=0.2$ and $0.5$. Several notable features are as follows.

\textit{Patterns in surface map.---} For $h=-0.5$, the $j_z$ and $B_z$ polarities become better aligned as $q$ increases (Figure~\ref{f:jz}). This is an interesting feature of the \citetalias{fan2003} flux tube, where $\vec{j}_\mathrm{P}$ and $\vec{B}_\mathrm{P}$ depend differently on $q$. The predominant poloidal current contribution $\vec{j}_\mathrm{P}\cdot\uvec{z}$ is independent of $q$ (Equation~(\ref{eq:jtpz})), whereas the toroidal contribution $\vec{j}_\mathrm{T}\cdot\uvec{z}$ merely appends a curved tail to the $j_z$ ribbon, creating a ``Yin-Yang'' pattern. Therefore, the $j_z$ patterns remain almost unchanged with $q$. Contrarily, the poloidal field contribution $\vec{B}_\mathrm{P}\cdot\uvec{z}$ linearly scales with $q$ (Equation~(\ref{eq:btpz})). As $q$ increases, the $B_z$ magnetic tongues rotate clockwise to better match the current ribbons. The PIL becomes largely parallel to the $j_z$ ribbons for $q=0.8$; the horizontal field $\vec{B}_h$ is also more sheared. As $h$ increases further, \texttt{U}-shaped field lines in the lower portion of the tube will graze the surface to create ``bald patches'', where $\vec{B}_h$ on the PIL directs from the negative to the positive side \citep{titov1993}.

\textit{Non-neutralized current vs. twist.---} The maximum $I_z$ and $R_z$ both positively correlate with $q$ (Figure~\ref{f:profile}(a)-(b)). This is because the $B_z$ and $j_z$ polarities become better aligned at higher $q$: the area occupied by DC increases. The $R_z$ values appear unexpectedly low for $q=0.8$ early on ($h < -0.8$). We find that the toroidal contribution in $I_z$ is initially negative (Figure~\ref{f:profile}(c)). This RC component is larger for $q=0.8$, leading to an overall smaller $R_z$. It is soon offset by the faster increasing DC component in the poloidal current.

\textit{Evolution of current and magnetic flux.---} As $h$ increases, $I_z$ and $R_z$ initially increase along with the surface magnetic flux $\Phi$ (Figure~\ref{f:profile}(d)). The former peak at $h\approx-R$, when a significant portion of the toroidal magnetic flux is still below the surface. As emergence proceeds, $I_z$ and $R_z$ will decrease due to the toroidal contribution, whereas $\Phi$ will continue to increase. For an untwisted flux tube, the maximum $\Phi$ is expected to occur at $h=-(R-3a)=-0.14$, when the crest of the tube fully emerges. Such is also the case for both $q=0.2$ and $0.5$. For $q=0.8$, however, $\Phi$ peaks significant earlier at $h=-0.37$ with a maximum 5\% higher than the total toroidal flux. This is owing to the strong poloidal field contribution (Equation~(\ref{eq:btpz})). A portion of highly twisted magnetic field lines will intersect the $z=0$ plane more than two times (appendix).

We conclude that the observed, non-neutralized vertical current (large $I_z$ and $R_z$) can arise from a neutralized flux tube (zero $I_\mathrm{T}$ and unity $R_\mathrm{T}$) due to the projected poloidal component. The conclusion holds for the early stages of emergence only, when the crest of the tube is partially emerged, and the axis obliquely intersects the surface. For the later stages, the tube axis becomes more perpendicular to the surface. The observed current becomes more neutralized as the toroidal component takes over.


\section{Discussion}
\label{sec:discussion}

For an idealized, torus-shaped flux tubes, as long as $B_\mathrm{T}$ decreases away from the tube axis, the surface $\vec{j}_\mathrm{P}\cdot\uvec{z}$ distribution is expected to be bipolar like \citetalias{fan2003}. Whether the surface current is neutralized will depend on the patterns of $\vec{B}_\mathrm{P}\cdot\uvec{z}$, which is sensitive to the global magnetic twist (Figure~\ref{f:jz}). Other conditions being equal, a flux tube with higher twist should possess higher magnetic helicity and free energy, which are known to drive solar eruptions. This, combined with the positive correlation between $R_z$ and $q$, provides a natural explanation to the increased eruptivity in ARs with significant non-neutralized currents.

We temper our conclusion with some caveats. MHD simulations have shown that it is difficult for a twisted flux tube to rise bodily into the corona \citep[e.g.,][]{fan2001,archontis2010}. The concave portions of the field lines are loaded with dense plasma, which prevents the tube axis from rising for more than a few scale heights above the surface. Indeed, the tongue-shaped $B_z$ and $j_z$ polarities (Figures~\ref{f:mapsb}(a)) frequently appear in observations \citep[e.g.,][]{poisson2016} without developing further into two circular polarities. If so, the $h \le -R$ phase in our model should be more relevant than $h > -R$ in realistic settings. In these early stages of flux emergence, the poloidal component is expected to play a prominent role. We note that counterexamples also exist where the axial flux fully emerges in simulation \citep[e.g.,][]{hood2009}.

In observations, CME-productive ARs frequently exhibit long-lived, narrow DC ribbon pairs that are well aligned with the PIL similar to the $q=0.8$ case (see, e.g., Figure 1 of \citealp{liuy2017} and Figure 3 of \citealp{avallone2020}). The overall patterns of $j_z$, however, are filamentary except for the DC ribbons (if present); evidence for a single shell of RC surrounding the DC is also lacking. The observed $R_z$ evolution sometimes show a fast increase with $\Phi$ and peaks earlier than $\Phi$ (see, e.g., Figure 2 of \citealp{vemareddy2019} and Figure 3 of \citealp{avallone2020}). Their absolute values, however, are considerably lower than in our model (Figure~\ref{f:profile}(b)). The maximum $R_z$ is typically $\sim$2 for CME-productive ARs, and at best reaches $\sim$4 when the central flux rope structure is largely isolated \citep{liuy2017}. 

It is worth mentioning that in ideal MHD, the evolution of $\vec{j}$ is fundamentally different from that of $\vec{B}$. There is no ``frozen-in'' theorem for $\vec{j}$ so $I$ is not conserved in a flux tube; $\partial \vec{j} / \partial t$ will depend not only on the velocity field $\vec{v}$ and $\vec{j}$ itself, but $\vec{B}$ as well \citep[e.g.,][]{inverarity1997}. Predicting the behavior of the electric current can be rather difficult without solving the MHD equations for the full system~\citep{parker1996a}.

Our simplistic approach neglects a rich variety of physics during flux emergence. We discuss several possible effects below.

\begin{itemize}[noitemsep,topsep=0pt,parsep=2pt]

\item As the flux tube approaches the surface, it will be flattened into a sheet-like structure due to the diminishing pressure scale height at the top of the convection zone \citep{spruit1987,Cheung:2008}. The internal $\vec{B}$ and $\vec{j}$ distributions will change.

\item The drastic expansion of the emerged flux due to magnetic buoyancy instabilities~\citep[section 3.3 of][]{cheung2014} can shunt the RC to the interface layer, as demonstrated by \citet{longcope2000} and \citet{torok2014}.

\item The expansion will also induce a Lorentz force, which drives twist or shear flows on the surface \citep{longcope2000,manchester2004,Fang:2012b}. These flows are capable of producing non-neutralized currents if they act close to the PIL \citep{dalmasse2015}.

\item The emerged flux may undergo internal or external reconnection to produce a new flux rope in the corona \citep{Archontis:2008,leake2013,Takasao:2015,Syntelis:2019,Toriumi:2019,Toriumi:2020}. This evolution in the magnetic field is reflected in changes of the current distribution.

\item Interactions between different emerging flux regions \citep[e.g.][]{Chintzoglou:2019}, and between existing sunspots and newly emerged regions \citep{Fan:2016,CheungRempel:2019}, are known to cause flares. The foot prints of the reconnecting quasi-separatrix layers \citep[QSLs; e.g.,][]{demoulin1996}, demarcated by flare ribbons, often overlap with the pre-flare current ribbons. Observations haven shown that the local $j_z$ can greatly enhance during the eruption \citep{janvier2014}.

\item The buffeting effect from the convective flows may fragment the magnetic flux~\citep{Cheung:2008,MartinezSykora:2008,SteinNordlund:2012,Chen:2017} and contribute to the filamentary structure of $j_z$~\citep{Yelles:2009,Fang:2012a}.

\end{itemize}

Due to these effects, the observed surface $\vec{B}$ and $\vec{j}$ are unlikely to be identical to those on a subsurface horizontal slice. A distinction between poloidal and toroidal components with respect to their subsurface origin can be difficult in practice. Nevertheless, the simple model here is capable of reproducing several key observational features. The projection effect may be important to the photospheric current distribution, in particular during the early stages of flux emergence.


\appendix

In \citetalias{fan2003}, the flux tube is kinematically lifted from subsurface ($z<0$) at a constant velocity $\vec{v}_0=v_0\uvec{z}$. As the tube intersects the $z=0$ plane, its magnetic field $\vec{B}_0=\vec{B}(z=0)$ and $\vec{v}_0$ induce an ideal electric field $\vec{E}\propto-\vec{v_0}\times\vec{B}_0$, which is used as an evolving boundary condition for the domain of interest ($z>0$). The initial potential field in the $z>0$ volume interacts with the newly emerged flux, resulting in a sigmoidal flux rope that eventually erupts.

For simplicity, we place the origin at $\vec{r}_0=(0,0,0)$ in the Cartesian coordinate. A horizontal slice at $z_0>0$ represents the photospheric condition when the tube is centered at a depth of $h=-z_0$ (Figure~\ref{f:mfr}).

We define two auxiliary variables: $\rho$ for the projected distance of a point $\vec{r}$ to the origin on the $y=0$ plane, and $w$ for the distance to the tube axis:
\begin{equation}
\label{eq:aux}
\begin{split}
\rho & = \left(x^2 +z^2 \right)^{1/2}, \\
w & = \left[( \rho - R )^2 + y^2 \right]^{1/2}. \\
\end{split}
\end{equation}

In our convention, both the Cartesian and spherical coordinate systems are right-handed (Figure~\ref{f:mfr}). The coordinates $(x, y, z)$ and $(r, \theta, \phi)$ are related as
\begin{equation}
\label{eq:car2sp}
\begin{split}
x & = r \sin{\theta} \sin{\phi}, \\
y & = r \cos{\theta}, \\
z & = r \sin{\theta} \cos{\phi}. \\
\end{split}
\end{equation}
Their unit vectors $(\uvec{x}, \uvec{y}, \uvec{z})$ and $(\uvec{r}, \uvec{\theta}, \uvec{\phi})$ are related as
\begin{equation}
\label{eq:ecar2sph}
\begin{split}
\uvec{x} & = \sin{\theta} \sin{\phi} \, \uvec{r} + \cos{\theta} \sin{\phi} \, \uvec{\theta} + \cos{\phi} \, \uvec{\phi}, \\
\uvec{y} & = \cos{\theta} \, \uvec{r} - \sin{\theta} \, \uvec{\theta}, \\
\uvec{z} & = \sin{\theta} \cos{\phi} \, \uvec{r} + \cos{\theta} \cos{\phi} \, \uvec{\theta} - \sin{\phi} \, \uvec{\phi}. \\
\end{split}
\end{equation}

In the spherical coordinate, the three components of the magnetic field $\vec{B}$ are
\begin{equation}
\label{eq:bsph3}
\begin{split}
B_r & 
	= B_t \dfrac{q R \cos{\theta}}{r \sin{\theta}} \exp\left(-w^2/a^2\right), \\ 
B_\theta & 
	= B_t \dfrac{q (r - R \sin{\theta})}{r \sin{\theta}} \exp\left(-w^2/a^2\right), \\
B_\phi & = B_t \dfrac{a}{r \sin{\theta}} \exp\left(-w^2/a^2\right). \\
\end{split}
\end{equation}
The three Cartesian components are
\begin{equation}
\label{eq:bcar3}
\begin{split}
B_x 
	& = B_t \dfrac{q x y + a z}{\rho^2} \exp\left(-w^2/a^2\right), \\
B_y 
	& = B_t q \dfrac{R - \rho}{\rho} \exp\left(-w^2/a^2\right), \\
B_z 
	& = B_t \dfrac{q y z - a x}{\rho^2} \exp\left(-w^2/a^2\right).
\end{split}
\end{equation}

In the spherical coordinate, the three components of the current density $\vec{j}$ are
\begin{equation}
\label{eq:jsph3}
\begin{split}
j_r 
	& = \dfrac{2 B_t R \cos\theta}{a \rho} \exp\left(-w^2/a^2\right), \\
j_\theta 
	& = \dfrac{ 2 B_t (r - R \sin{\theta})}{a \rho} \exp\left(-w^2/a^2\right),  \\
j_\phi 
	& = - \dfrac{B_t q}{\rho} \left( \dfrac{2w^2}{a^2} - \dfrac{R}{\rho} - 1 \right) \exp\left(-w^2/a^2\right).
\end{split}
\end{equation}
The three Cartesian components are
\begin{equation}
\label{eq:jcar3}
\begin{split}
j_x 
	& = - \dfrac{B_t}{\rho^2} \left\{ q \left[ \dfrac{2 w^2}{a^2} - \dfrac{R}{\rho} - 1 \right] z - \dfrac{2 x y}{a} \right\} \exp\left(-w^2/a^2\right), \\
j_y 
	& = \dfrac{2 B_t (R-\rho)}{a \rho} \exp\left(-w^2/a^2\right) ,\\
j_z 
	& =  \dfrac{B_t}{\rho^2} \left\{ q \left[ \dfrac{2 w^2}{a^2} - \dfrac{R}{\rho} - 1 \right] x + \dfrac{2 y z}{a} \right\} \exp\left(-w^2/a^2\right). \\
\end{split}
\end{equation}

A torus coordinate $(w, \psi, \phi)$ is related to the Cartesian coordinate as
\begin{equation}
\label{eq:car2toi}
\begin{split}
x & = (R + w \cos{\psi}) \sin{\phi}, \\
y & = - w \sin{\psi}, \\
z & = (R + w \cos{\psi}) \cos{\phi}. \\
\end{split}
\end{equation}
Here $w$ is the radius with respect to the tube axis, and $\phi$ is the \textit{toroidal} angle with respect to the center of the torus. They are identically defined as in the spherical coordinate. The new coordinate $\psi$ defines the \textit{poloidal} angle, which is the polar angle with respect to the tube axis. It is 0 in the $x$-$z$ plane, and increase in the same direction as $\theta$ (Figure~\ref{f:mfr}).

The unit vectors $(\uvec{w}, \uvec{\psi})$ and $(\uvec{r}, \uvec{\theta})$ are related as
\begin{equation}
\label{eq:etoi2sph}
\begin{split}
\uvec{w} & = \dfrac{r^2 - R \rho}{wr} \, \uvec{r} - \dfrac{Ry}{wr} \, \uvec{\theta}, \\
\uvec{\psi} & = \dfrac{Ry}{wr} \, \uvec{r} + \dfrac{r^2 - R \rho}{wr} \, \uvec{\theta}. \\
\end{split}
\end{equation}

In this definition, the axial component is toroidal ($\uvec{\phi}$). The non-axial components include the radial ($\uvec{w}$) and poloidal ($\uvec{\psi}$) contributions. Using Equations~(\ref{eq:bsph3}), (\ref{eq:jsph3}), and (\ref{eq:etoi2sph}), we find
\begin{equation}
\label{eq:btoi3}
\begin{split}
B_w			& =  0,  \\
B_\psi	& =  \dfrac{q w B_t}{\rho} \exp\left(-w^2/a^2\right), \\
\end{split}
\end{equation}
\begin{equation}
\label{eq:jtoi3}
\begin{split}
j_w		& =  0,  \\
j_\psi		& =  \dfrac{2wB_t}{a\rho} \exp\left(-w^2/a^2\right). \\
\end{split}
\end{equation}
This indicates that the non-axial $\vec{B}$ and $\vec{j}$ are purely poloidal.

Using Equations~(\ref{eq:btp}), (\ref{eq:jtp}), (\ref{eq:ecar2sph}), (\ref{eq:bsph3}), and (\ref{eq:jsph3}), we can evaluate the contributions of poloidal and toroidal components in $B_z$ and $j_z$:
\begin{equation}
\label{eq:btpz}
\begin{split}
\vec{B}_\mathrm{P} \cdot \uvec{z} & = B_t \dfrac{q y z}{\rho^2} \exp\left(-w^2/a^2\right), \\
\vec{B}_\mathrm{T} \cdot \uvec{z} & = - B_t \dfrac{a x}{\rho^2} \exp\left(-w^2/a^2\right), \\
\end{split}
\end{equation}
\begin{equation}
\label{eq:jtpz}
\begin{split}
\vec{j}_\mathrm{P} \cdot \uvec{z} 
	& = \dfrac{2 B_t yz}{a \rho^2} \exp\left(-w^2/a^2\right), \\
\vec{j}_\mathrm{T} \cdot \uvec{z} 
	& = \dfrac{B_t q x}{\rho^2} \left( \dfrac{2w^2}{a^2} - \dfrac{R}{\rho} - 1 \right) \exp\left(-w^2/a^2\right). \\
\end{split}
\end{equation}


\begin{figure}[t!]
\centerline{\includegraphics[width=1.0\textwidth]{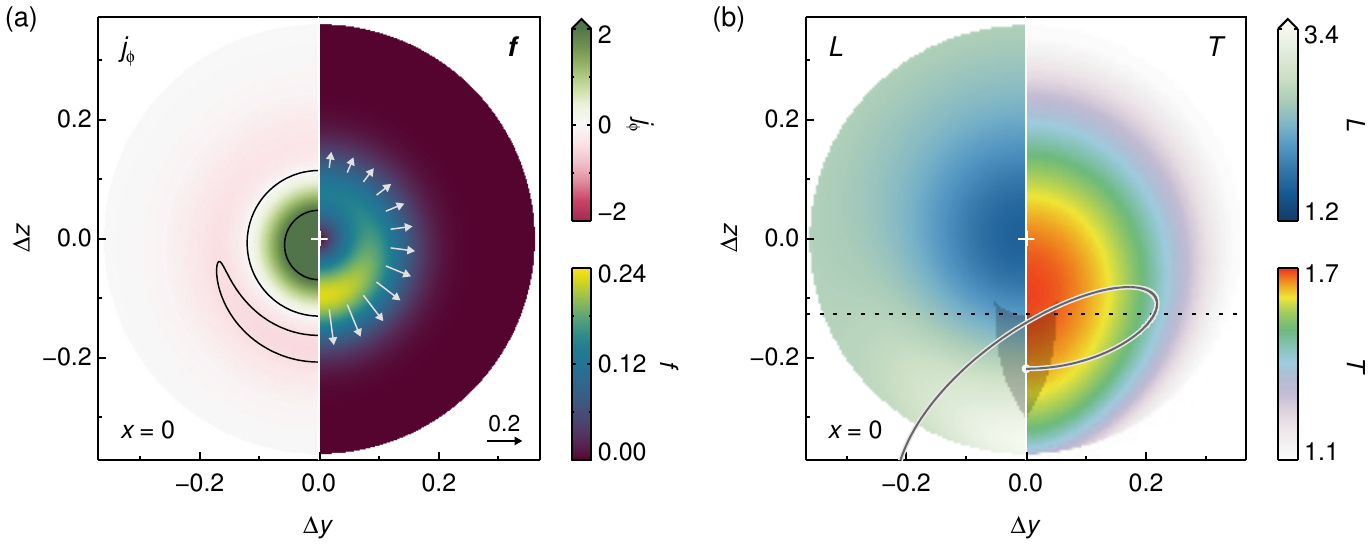}}
\caption{Properties of a \citetalias{fan2003} flux tube on a cross section ($q=0.8$, $\phi=x=0$). (a) Map of toroidal current density $j_\phi$ (left) and Lorentz force $\vec{f}$ (right). The contours are for $j_\phi = -0.45$, $0$, and $2$. The background colors show the force amplitude; the arrows show the force vectors at $w=0.12$. The tube axis is marked with a cross. (b) Map of the length $L$ (left) and the number of turns $T$ around the axis (right) for field lines passing through the $x=0$ plane and end at $\Delta z=-0.5$. Horizontal dotted line shows the photosphere for $h=-0.37$ when the observed magnetic flux reaches maximum. A portion of field lines will cross the photosphere four times; one such field line is shown (for $x>0$). The wedge-shaped shaded region denotes their intersections with $x=0$.}

\label{f:cutx0}
\end{figure}


The semi-torus tube is axisymmetric about the $y$-axis. However, on a cross section with constant $\phi$, variables are not axisymmetric about the tube axis.  For example, for $j_\phi$, the inner, lower portion of the RC sheath appears to be stronger than other parts (Figure~\ref{f:cutx0}(a)).

We evaluate the Lorentz force density $\vec{f}=\vec{j}\times\vec{B}$ on the cross section $\phi=0$. As both the radial field ($B_w$) and radial current density ($j_w$) are zero, $\vec{f}$ must be purely radial. The force points away from the axis so the tube tends to expand. It is the strongest at the lower portion of the DC core (Figure~\ref{f:cutx0}(a)).

We trace field lines in both directions from the $\phi=0$ plane to investigate their lengths $L$ and turns of twist $T$ around the axis (Figure~\ref{f:cutx0}(b)). The latter is evaluated from the change of $\psi$ in the torus coordinate. For $q=0.8$, the axis of the semi-torus has $L = \pi R = 1.57$; field lines passing through the lower edge are more than twice in length due to twist. Field lines close to the axis has $T= q R / 2 a = 1.67$; those passing slightly below the axis have a larger $T$ due to increased $L$.

For the $q=0.8$ case, we find that a portion of the field lines may intersect the photosphere four times (Figure 8(b)). These field lines cross the $\phi=0$ plane in a wedge-shaped region below the tube axis, where $L$ and $T$ are both large. This is a result of competition between the curvatures given by the flux tube axis and by the twist in the flux tube. No such field lines are present for the $q=0.2$ and $0.5$ cases.


\begin{acks}
XS is partially supported by NSF award 1848250 and NASA award 80NSSC190263. MCMC acknowledges support from NASA award 80NSSC19K0088. This work was initiated during a scientific visit to the Institute for Astronomy by MCMC with travel support from NSF award 1848250. We thank T. T\"{o}r\"{o}k, Y. Liu, and V. S. Titov for their valuable comments. Visualization softwares include the \texttt{Plotly} \textit{Python} package.
\end{acks}


%
%
\bibliographystyle{spr-mp-sola}
\bibliography{neutral}

\begin{thebibliography}{54}
\ifx\bisbn     \undefined \def\bisbn  #1{ISBN #1}\fi
\ifx\binits    \undefined \def\binits#1{#1}\fi
\ifx\bauthor   \undefined \def\bauthor#1{#1}\fi
\ifx\batitle   \undefined \def\batitle#1{#1}\fi
\ifx\bjtitle   \undefined \def\bjtitle#1{\textit{#1}}\fi
\ifx\bvolume   \undefined \def\bvolume#1{\textbf{#1}}\fi
\ifx\byear     \undefined \def\byear#1{#1}\fi
\ifx\bissue    \undefined \def\bissue#1{#1}\fi
\ifx\bfpage    \undefined \def\bfpage#1{#1}\fi
\ifx\blpage    \undefined \def\blpage #1{#1}\fi
\ifx\burl      \undefined \def\burl#1{#1}\fi
\ifx\href      \undefined \def\href#1#2{#2}\fi
\ifx\betal     \undefined \def\betal{et al.}\fi
\ifx\bctitle   \undefined \def\bctitle#1{#1}\fi
\ifx\beditor   \undefined \def\beditor#1{#1}\fi
\ifx\bbtitle   \undefined \def\bbtitle#1{\textit{#1}}\fi
\ifx\bedition  \undefined \def\bedition#1{#1}\fi
\ifx\bseriesno \undefined \def\bseriesno#1{\textbf{#1}}\fi
\ifx\blocation \undefined \def\blocation#1{#1}\fi
\ifx\bsertitle \undefined \def\bsertitle#1{\textit{#1}}\fi
\ifx\bsnm      \undefined \def\bsnm#1{#1}\fi
\ifx\bsuffix   \undefined \def\bsuffix#1{#1}\fi
\ifx\bparticle \undefined \def\bparticle#1{#1}\fi
\ifx\barticle  \undefined \def\barticle#1{}\fi
\ifx\binstitute  \undefined \def\binstitute#1{#1}\fi
\ifx\bpublisher  \undefined \def\bpublisher#1{#1}\fi
\ifx\doiurl    \undefined \def\doiurl#1{\href{#1}{DOI}}\fi
\makeatletter
\def\safeHref#1#2#3{\in@{http}{#2}\ifin@\href{#2}{#3}\else\href{#1#2}{#3}\fi}
\makeatother
\ifx\adsurl    \undefined
  \def\adsurl#1{\safeHref{https://ui.adsabs.harvard.edu/abs/}{#1}{ADS}}\fi
\ifx\arxivurl  \undefined
  \def\arxivurl#1{\safeHref{http://arxiv.org/abs/}{#1}{arXiv}}\fi
\ifx\botherref \undefined \def\botherref#1{}\fi
\ifx\url       \undefined \def\url#1{#1}\fi
\ifx\bchapter  \undefined \def\bchapter#1{}\fi
\ifx\bbook     \undefined \def\bbook#1{}\fi
\ifx\bcomment  \undefined \def\bcomment#1{#1}\fi
\ifx\oauthor   \undefined \def\oauthor#1{#1}\fi
\ifx\citeauthoryear \undefined\def \citeauthoryear#1{#1}\fi
\def\endbibitem {}
\ifx\bconflocation  \undefined \def\bconflocation#1{#1} \fi

\bibitem[\protect\citeauthoryear{{Archontis} and {Hood}}{2010}]{archontis2010}
\begin{barticle}
\bauthor{\bsnm{{Archontis}}, \binits{V.}},
\bauthor{\bsnm{{Hood}}, \binits{A.W.}}:
\byear{2010},
\batitle{{Flux emergence and coronal eruption}}.
\bjtitle{\aap}
\bvolume{514},
\bfpage{A56}.
\doiurl{https://doi.org/10.1051/0004-6361/200913502}.
\adsurl{2010A&A...514A..56A}.
\end{barticle}
\endbibitem

\bibitem[\protect\citeauthoryear{{Archontis} and
  {T{\"o}r{\"o}k}}{2008}]{Archontis:2008}
\begin{barticle}
\bauthor{\bsnm{{Archontis}}, \binits{V.}},
\bauthor{\bsnm{{T{\"o}r{\"o}k}}, \binits{T.}}:
\byear{2008},
\batitle{{Eruption of magnetic flux ropes during flux emergence}}.
\bjtitle{\aap}
\bvolume{492},
\bfpage{L35}.
\doiurl{https://doi.org/10.1051/0004-6361:200811131}.
\adsurl{2008A&A...492L..35A}.
\end{barticle}
\endbibitem

\bibitem[\protect\citeauthoryear{{Avallone} and {Sun}}{2020}]{avallone2020}
\begin{barticle}
\bauthor{\bsnm{{Avallone}}, \binits{E.A.}},
\bauthor{\bsnm{{Sun}}, \binits{X.}}:
\byear{2020},
\batitle{{Electric Current Neutralization in Solar Active Regions and Its
  Relation to Eruptive Activity}}.
\bjtitle{\apj}
\bvolume{893},
\bfpage{123}.
\doiurl{https://doi.org/10.3847/1538-4357/ab7afa}.
\adsurl{2020ApJ...893..123A}.
\end{barticle}
\endbibitem

\bibitem[\protect\citeauthoryear{{Chen}, {Rempel}, and {Fan}}{2017}]{Chen:2017}
\begin{barticle}
\bauthor{\bsnm{{Chen}}, \binits{F.}},
\bauthor{\bsnm{{Rempel}}, \binits{M.}},
\bauthor{\bsnm{{Fan}}, \binits{Y.}}:
\byear{2017},
\batitle{{Emergence of Magnetic Flux Generated in a Solar Convective Dynamo. I.
  The Formation of Sunspots and Active Regions, and The Origin of Their
  Asymmetries}}.
\bjtitle{\apj}
\bvolume{846},
\bfpage{149}.
\doiurl{https://doi.org/10.3847/1538-4357/aa85a0}.
\adsurl{2017ApJ...846..149C}.
\end{barticle}
\endbibitem

\bibitem[\protect\citeauthoryear{{Cheung} and {Isobe}}{2014}]{cheung2014}
\begin{barticle}
\bauthor{\bsnm{{Cheung}}, \binits{M.C.M.}},
\bauthor{\bsnm{{Isobe}}, \binits{H.}}:
\byear{2014},
\batitle{{Flux Emergence (Theory)}}.
\bjtitle{Living Rev. Sol. Phys.}
\bvolume{11},
\bfpage{3}.
\doiurl{https://doi.org/10.12942/lrsp-2014-3}.
\adsurl{2014LRSP...11....3C}.
\end{barticle}
\endbibitem

\bibitem[\protect\citeauthoryear{{Cheung} et~al.}{2008}]{Cheung:2008}
\begin{barticle}
\bauthor{\bsnm{{Cheung}}, \binits{M.C.M.}},
\bauthor{\bsnm{{Sch{\"u}ssler}}, \binits{M.}},
\bauthor{\bsnm{{Tarbell}}, \binits{T.D.}},
\bauthor{\bsnm{{Title}}, \binits{A.M.}}:
\byear{2008},
\batitle{{Solar Surface Emerging Flux Regions: A Comparative Study of Radiative
  MHD Modeling and Hinode SOT Observations}}.
\bjtitle{\apj}
\bvolume{687},
\bfpage{1373}.
\doiurl{https://doi.org/10.1086/591245}.
\adsurl{2008ApJ...687.1373C}.
\end{barticle}
\endbibitem

\bibitem[\protect\citeauthoryear{{Cheung} et~al.}{2019}]{CheungRempel:2019}
\begin{barticle}
\bauthor{\bsnm{{Cheung}}, \binits{M.C.M.}},
\bauthor{\bsnm{{Rempel}}, \binits{M.}},
\bauthor{\bsnm{{Chintzoglou}}, \binits{G.}},
\bauthor{\bsnm{{Chen}}, \binits{F.}},
\bauthor{\bsnm{{Testa}}, \binits{P.}},
\bauthor{\bsnm{{Mart{\'\i}nez-Sykora}}, \binits{J.}},
\bauthor{\bsnm{{Sainz Dalda}}, \binits{A.}},
\bauthor{\bsnm{{DeRosa}}, \binits{M.L.}},
\bauthor{\bsnm{{Malanushenko}}, \binits{A.}},
\bauthor{\bsnm{{Hansteen}}, \binits{V.}},
\bauthor{\bsnm{{De Pontieu}}, \binits{B.}},
\bauthor{\bsnm{{Carlsson}}, \binits{M.}},
\bauthor{\bsnm{{Gudiksen}}, \binits{B.}},
\bauthor{\bsnm{{McIntosh}}, \binits{S.W.}}:
\byear{2019},
\batitle{{A comprehensive three-dimensional radiative magnetohydrodynamic
  simulation of a solar flare}}.
\bjtitle{Nature Astronomy}
\bvolume{3},
\bfpage{160}.
\doiurl{https://doi.org/10.1038/s41550-018-0629-3}.
\adsurl{2019NatAs...3..160C}.
\end{barticle}
\endbibitem

\bibitem[\protect\citeauthoryear{{Chintzoglou} et~al.}{2019}]{Chintzoglou:2019}
\begin{barticle}
\bauthor{\bsnm{{Chintzoglou}}, \binits{G.}},
\bauthor{\bsnm{{Zhang}}, \binits{J.}},
\bauthor{\bsnm{{Cheung}}, \binits{M.C.M.}},
\bauthor{\bsnm{{Kazachenko}}, \binits{M.}}:
\byear{2019},
\batitle{{The Origin of Major Solar Activity: Collisional Shearing between
  Nonconjugated Polarities of Multiple Bipoles Emerging within Active
  Regions}}.
\bjtitle{\apj}
\bvolume{871},
\bfpage{67}.
\doiurl{https://doi.org/10.3847/1538-4357/aaef30}.
\adsurl{2019ApJ...871...67C}.
\end{barticle}
\endbibitem

\bibitem[\protect\citeauthoryear{{Dalmasse} et~al.}{2015}]{dalmasse2015}
\begin{barticle}
\bauthor{\bsnm{{Dalmasse}}, \binits{K.}},
\bauthor{\bsnm{{Aulanier}}, \binits{G.}},
\bauthor{\bsnm{{D{\'e}moulin}}, \binits{P.}},
\bauthor{\bsnm{{Kliem}}, \binits{B.}},
\bauthor{\bsnm{{T{\"o}r{\"o}k}}, \binits{T.}},
\bauthor{\bsnm{{Pariat}}, \binits{E.}}:
\byear{2015},
\batitle{{The Origin of Net Electric Currents in Solar Active Regions}}.
\bjtitle{\apj}
\bvolume{810},
\bfpage{17}.
\doiurl{https://doi.org/10.1088/0004-637X/810/1/17}.
\adsurl{2015ApJ...810...17D}.
\end{barticle}
\endbibitem

\bibitem[\protect\citeauthoryear{{D{\'e}moulin}, {Priest}, and
  {Lonie}}{1996}]{demoulin1996}
\begin{barticle}
\bauthor{\bsnm{{D{\'e}moulin}}, \binits{P.}},
\bauthor{\bsnm{{Priest}}, \binits{E.R.}},
\bauthor{\bsnm{{Lonie}}, \binits{D.P.}}:
\byear{1996},
\batitle{{Three-dimensional magnetic reconnection without null points 2.
  Application to twisted flux tubes}}.
\bjtitle{JGR}
\bvolume{101},
\bfpage{7631}.
\doiurl{https://doi.org/10.1029/95JA03558}.
\adsurl{1996JGR...101.7631D}.
\end{barticle}
\endbibitem

\bibitem[\protect\citeauthoryear{{Falconer}}{2001}]{falconer2001}
\begin{barticle}
\bauthor{\bsnm{{Falconer}}, \binits{D.A.}}:
\byear{2001},
\batitle{{A prospective method for predicting coronal mass ejections from
  vector magnetograms}}.
\bjtitle{\jgr}
\bvolume{106},
\bfpage{25185}.
\doiurl{https://doi.org/10.1029/2000JA004005}.
\adsurl{2001JGR...10625185F}.
\end{barticle}
\endbibitem

\bibitem[\protect\citeauthoryear{{Fan}}{2001}]{fan2001}
\begin{barticle}
\bauthor{\bsnm{{Fan}}, \binits{Y.}}:
\byear{2001},
\batitle{{The Emergence of a Twisted {\ensuremath{\Omega}}-Tube into the Solar
  Atmosphere}}.
\bjtitle{\apjl}
\bvolume{554},
\bfpage{L111}.
\doiurl{https://doi.org/10.1086/320935}.
\adsurl{2001ApJ...554L.111F}.
\end{barticle}
\endbibitem

\bibitem[\protect\citeauthoryear{{Fan}}{2009}]{fan2009lrsp}
\begin{barticle}
\bauthor{\bsnm{{Fan}}, \binits{Y.}}:
\byear{2009},
\batitle{{Magnetic Fields in the Solar Convection Zone}}.
\bjtitle{Living Reviews in Solar Physics}
\bvolume{6},
\bfpage{4}.
\doiurl{https://doi.org/10.12942/lrsp-2009-4}.
\adsurl{2009LRSP....6....4F}.
\end{barticle}
\endbibitem

\bibitem[\protect\citeauthoryear{{Fan}}{2016}]{Fan:2016}
\begin{barticle}
\bauthor{\bsnm{{Fan}}, \binits{Y.}}:
\byear{2016},
\batitle{{Modeling the Initiation of the 2006 December 13 Coronal Mass Ejection
  in AR 10930: The Structure and Dynamics of the Erupting Flux Rope}}.
\bjtitle{\apj}
\bvolume{824},
\bfpage{93}.
\doiurl{https://doi.org/10.3847/0004-637X/824/2/93}.
\adsurl{2016ApJ...824...93F}.
\end{barticle}
\endbibitem

\bibitem[\protect\citeauthoryear{{Fan} and {Gibson}}{2003}]{fan2003}
\begin{barticle}
\bauthor{\bsnm{{Fan}}, \binits{Y.}},
\bauthor{\bsnm{{Gibson}}, \binits{S.E.}}:
\byear{2003},
\batitle{{The Emergence of a Twisted Magnetic Flux Tube into a Preexisting
  Coronal Arcade}}.
\bjtitle{\apjl}
\bvolume{589},
\bfpage{L105}.
\doiurl{https://doi.org/10.1086/375834}.
\adsurl{2003ApJ...589L.105F}.
\end{barticle}
\endbibitem

\bibitem[\protect\citeauthoryear{{Fang} et~al.}{2012a}]{Fang:2012b}
\begin{barticle}
\bauthor{\bsnm{{Fang}}, \binits{F.}},
\bauthor{\bsnm{{Manchester}}, \binits{I.} \bsuffix{Ward}},
\bauthor{\bsnm{{Abbett}}, \binits{W.P.}},
\bauthor{\bsnm{{van der Holst}}, \binits{B.}}:
\byear{2012}a,
\batitle{{Buildup of Magnetic Shear and Free Energy during Flux Emergence and
  Cancellation}}.
\bjtitle{\apj}
\bvolume{754},
\bfpage{15}.
\doiurl{https://doi.org/10.1088/0004-637X/754/1/15}.
\adsurl{2012ApJ...754...15F}.
\end{barticle}
\endbibitem

\bibitem[\protect\citeauthoryear{{Fang} et~al.}{2012b}]{Fang:2012a}
\begin{barticle}
\bauthor{\bsnm{{Fang}}, \binits{F.}},
\bauthor{\bsnm{{Manchester}}, \binits{I.} \bsuffix{Ward}},
\bauthor{\bsnm{{Abbett}}, \binits{W.P.}},
\bauthor{\bsnm{{van der Holst}}, \binits{B.}}:
\byear{2012}b,
\batitle{{Dynamic Coupling of Convective Flows and Magnetic Field during Flux
  Emergence}}.
\bjtitle{\apj}
\bvolume{745},
\bfpage{37}.
\doiurl{https://doi.org/10.1088/0004-637X/745/1/37}.
\adsurl{2012ApJ...745...37F}.
\end{barticle}
\endbibitem

\bibitem[\protect\citeauthoryear{{Forbes}}{2010}]{forbes2010}
\begin{bbook}
\bauthor{\bsnm{{Forbes}}, \binits{T.}}:
\byear{2010},
In: \beditor{\bsnm{{Schrijver}}, \binits{C.J.}},
\beditor{\bsnm{{Siscoe}}, \binits{G.L.}} (eds.)
\bbtitle{{Models of coronal mass ejections and flares}},
\bfpage{159}.
\adsurl{2010hssr.book..159F}.
\end{bbook}
\endbibitem

\bibitem[\protect\citeauthoryear{{Gary} and {Rabin}}{1995}]{gary1995}
\begin{barticle}
\bauthor{\bsnm{{Gary}}, \binits{G.A.}},
\bauthor{\bsnm{{Rabin}}, \binits{D.}}:
\byear{1995},
\batitle{{Line-of-Sight Magnetic Flux Imbalances Caused by Electric Currents}}.
\bjtitle{\solphys}
\bvolume{157},
\bfpage{185}.
\doiurl{https://doi.org/10.1007/BF00680616}.
\adsurl{1995SoPh..157..185G}.
\end{barticle}
\endbibitem

\bibitem[\protect\citeauthoryear{{Georgoulis}, {Titov}, and
  {Miki{\'c}}}{2012}]{georgoulis2012}
\begin{barticle}
\bauthor{\bsnm{{Georgoulis}}, \binits{M.K.}},
\bauthor{\bsnm{{Titov}}, \binits{V.S.}},
\bauthor{\bsnm{{Miki{\'c}}}, \binits{Z.}}:
\byear{2012},
\batitle{{Non-neutralized Electric Current Patterns in Solar Active Regions:
  Origin of the Shear-generating Lorentz Force}}.
\bjtitle{\apjl}
\bvolume{761},
\bfpage{61}.
\doiurl{https://doi.org/10.1088/0004-637X/761/1/61}.
\adsurl{2012ApJ...761...61G}.
\end{barticle}
\endbibitem

\bibitem[\protect\citeauthoryear{{Gibson} et~al.}{2004}]{gibson2004}
\begin{barticle}
\bauthor{\bsnm{{Gibson}}, \binits{S.E.}},
\bauthor{\bsnm{{Fan}}, \binits{Y.}},
\bauthor{\bsnm{{Mandrini}}, \binits{C.}},
\bauthor{\bsnm{{Fisher}}, \binits{G.}},
\bauthor{\bsnm{{Demoulin}}, \binits{P.}}:
\byear{2004},
\batitle{{Observational Consequences of a Magnetic Flux Rope Emerging into the
  Corona}}.
\bjtitle{\apj}
\bvolume{617},
\bfpage{600}.
\doiurl{https://doi.org/10.1086/425294}.
\adsurl{2004ApJ...617..600G}.
\end{barticle}
\endbibitem

\bibitem[\protect\citeauthoryear{{Gosain}, {D{\'e}moulin}, and {L{\'o}pez
  Fuentes}}{2014}]{gosain2014}
\begin{barticle}
\bauthor{\bsnm{{Gosain}}, \binits{S.}},
\bauthor{\bsnm{{D{\'e}moulin}}, \binits{P.}},
\bauthor{\bsnm{{L{\'o}pez Fuentes}}, \binits{M.}}:
\byear{2014},
\batitle{{Distribution of Electric Currents in Sunspots from Photosphere to
  Corona}}.
\bjtitle{\apj}
\bvolume{793},
\bfpage{15}.
\doiurl{https://doi.org/10.1088/0004-637X/793/1/15}.
\adsurl{2014ApJ...793...15G}.
\end{barticle}
\endbibitem

\bibitem[\protect\citeauthoryear{{Hood} et~al.}{2009}]{hood2009}
\begin{barticle}
\bauthor{\bsnm{{Hood}}, \binits{A.W.}},
\bauthor{\bsnm{{Archontis}}, \binits{V.}},
\bauthor{\bsnm{{Galsgaard}}, \binits{K.}},
\bauthor{\bsnm{{Moreno-Insertis}}, \binits{F.}}:
\byear{2009},
\batitle{{The emergence of toroidal flux tubes from beneath the solar
  photosphere}}.
\bjtitle{\aap}
\bvolume{503},
\bfpage{999}.
\doiurl{https://doi.org/10.1051/0004-6361/200912189}.
\adsurl{2009A&A...503..999H}.
\end{barticle}
\endbibitem

\bibitem[\protect\citeauthoryear{{Inverarity} and
  {Titov}}{1997}]{inverarity1997}
\begin{barticle}
\bauthor{\bsnm{{Inverarity}}, \binits{G.W.}},
\bauthor{\bsnm{{Titov}}, \binits{V.S.}}:
\byear{1997},
\batitle{{Formation of current layers in three-dimensional, inhomogeneous
  coronal magnetic fields by photospheric motions}}.
\bjtitle{\jgr}
\bvolume{102},
\bfpage{22285}.
\doiurl{https://doi.org/10.1029/97JA01674}.
\adsurl{1997JGR...10222285I}.
\end{barticle}
\endbibitem

\bibitem[\protect\citeauthoryear{{Janvier} et~al.}{2014}]{janvier2014}
\begin{barticle}
\bauthor{\bsnm{{Janvier}}, \binits{M.}},
\bauthor{\bsnm{{Aulanier}}, \binits{G.}},
\bauthor{\bsnm{{Bommier}}, \binits{V.}},
\bauthor{\bsnm{{Schmieder}}, \binits{B.}},
\bauthor{\bsnm{{D{\'e}moulin}}, \binits{P.}},
\bauthor{\bsnm{{Pariat}}, \binits{E.}}:
\byear{2014},
\batitle{{Electric Currents in Flare Ribbons: Observations and
  Three-dimensional Standard Model}}.
\bjtitle{\apj}
\bvolume{788},
\bfpage{60}.
\doiurl{https://doi.org/10.1088/0004-637X/788/1/60}.
\adsurl{2014ApJ...788...60J}.
\end{barticle}
\endbibitem

\bibitem[\protect\citeauthoryear{{Kontogiannis}
  et~al.}{2019}]{kontogiannis2019}
\begin{barticle}
\bauthor{\bsnm{{Kontogiannis}}, \binits{I.}},
\bauthor{\bsnm{{Georgoulis}}, \binits{M.K.}},
\bauthor{\bsnm{{Guerra}}, \binits{J.A.}},
\bauthor{\bsnm{{Park}}, \binits{S.-H.}},
\bauthor{\bsnm{{Bloomfield}}, \binits{D.S.}}:
\byear{2019},
\batitle{{Which Photospheric Characteristics Are Most Relevant to Active-Region
  Coronal Mass Ejections?}}
\bjtitle{\solphys}
\bvolume{294},
\bfpage{130}.
\doiurl{https://doi.org/10.1007/s11207-019-1523-6}.
\adsurl{2019SoPh..294..130K}.
\end{barticle}
\endbibitem

\bibitem[\protect\citeauthoryear{{Leake}, {Linton}, and
  {T{\"o}r{\"o}k}}{2013}]{leake2013}
\begin{barticle}
\bauthor{\bsnm{{Leake}}, \binits{J.E.}},
\bauthor{\bsnm{{Linton}}, \binits{M.G.}},
\bauthor{\bsnm{{T{\"o}r{\"o}k}}, \binits{T.}}:
\byear{2013},
\batitle{{Simulations of Emerging Magnetic Flux. I. The Formation of Stable
  Coronal Flux Ropes}}.
\bjtitle{\apj}
\bvolume{778},
\bfpage{99}.
\doiurl{https://doi.org/10.1088/0004-637X/778/2/99}.
\adsurl{2013ApJ...778...99L}.
\end{barticle}
\endbibitem

\bibitem[\protect\citeauthoryear{{Leka} et~al.}{1996}]{leka1996}
\begin{barticle}
\bauthor{\bsnm{{Leka}}, \binits{K.D.}},
\bauthor{\bsnm{{Canfield}}, \binits{R.C.}},
\bauthor{\bsnm{{McClymont}}, \binits{A.N.}},
\bauthor{\bsnm{{van Driel-Gesztelyi}}, \binits{L.}}:
\byear{1996},
\batitle{{Evidence for Current-carrying Emerging Flux}}.
\bjtitle{\apj}
\bvolume{462},
\bfpage{547}.
\doiurl{https://doi.org/10.1086/177171}.
\adsurl{1996ApJ...462..547L}.
\end{barticle}
\endbibitem

\bibitem[\protect\citeauthoryear{{Lites} et~al.}{1995}]{lites1995}
\begin{barticle}
\bauthor{\bsnm{{Lites}}, \binits{B.W.}},
\bauthor{\bsnm{{Low}}, \binits{B.C.}},
\bauthor{\bsnm{{Martinez Pillet}}, \binits{V.}},
\bauthor{\bsnm{{Seagraves}}, \binits{P.}},
\bauthor{\bsnm{{Skumanich}}, \binits{A.}},
\bauthor{\bsnm{{Frank}}, \binits{Z.A.}},
\bauthor{\bsnm{{Shine}}, \binits{R.A.}},
\bauthor{\bsnm{{Tsuneta}}, \binits{S.}}:
\byear{1995},
\batitle{{The Possible Ascent of a Closed Magnetic System through the
  Photosphere}}.
\bjtitle{\apj}
\bvolume{446},
\bfpage{877}.
\doiurl{https://doi.org/10.1086/175845}.
\adsurl{1995ApJ...446..877L}.
\end{barticle}
\endbibitem

\bibitem[\protect\citeauthoryear{{Liu} et~al.}{2017}]{liuy2017}
\begin{barticle}
\bauthor{\bsnm{{Liu}}, \binits{Y.}},
\bauthor{\bsnm{{Sun}}, \binits{X.}},
\bauthor{\bsnm{{T{\"o}r{\"o}k}}, \binits{T.}},
\bauthor{\bsnm{{Titov}}, \binits{V.S.}},
\bauthor{\bsnm{{Leake}}, \binits{J.E.}}:
\byear{2017},
\batitle{{Electric-current Neutralization, Magnetic Shear, and Eruptive
  Activity in Solar Active Regions}}.
\bjtitle{\apjl}
\bvolume{846},
\bfpage{L6}.
\doiurl{https://doi.org/10.3847/2041-8213/aa861e}.
\adsurl{2017ApJ...846L...6L}.
\end{barticle}
\endbibitem

\bibitem[\protect\citeauthoryear{{Longcope} and {Welsch}}{2000}]{longcope2000}
\begin{barticle}
\bauthor{\bsnm{{Longcope}}, \binits{D.W.}},
\bauthor{\bsnm{{Welsch}}, \binits{B.T.}}:
\byear{2000},
\batitle{{A Model for the Emergence of a Twisted Magnetic Flux Tube}}.
\bjtitle{\apj}
\bvolume{545},
\bfpage{1089}.
\doiurl{https://doi.org/10.1086/317846}.
\adsurl{2000ApJ...545.1089L}.
\end{barticle}
\endbibitem

\bibitem[\protect\citeauthoryear{{L{\'o}pez Fuentes}
  et~al.}{2000}]{lopezfuentes2000}
\begin{barticle}
\bauthor{\bsnm{{L{\'o}pez Fuentes}}, \binits{M.C.}},
\bauthor{\bsnm{{Demoulin}}, \binits{P.}},
\bauthor{\bsnm{{Mandrini}}, \binits{C.H.}},
\bauthor{\bsnm{{van Driel-Gesztelyi}}, \binits{L.}}:
\byear{2000},
\batitle{{The Counterkink Rotation of a Non-Hale Active Region}}.
\bjtitle{\apj}
\bvolume{544},
\bfpage{540}.
\doiurl{https://doi.org/10.1086/317180}.
\adsurl{2000ApJ...544..540L}.
\end{barticle}
\endbibitem

\bibitem[\protect\citeauthoryear{{Luoni} et~al.}{2011}]{luoni2011}
\begin{barticle}
\bauthor{\bsnm{{Luoni}}, \binits{M.L.}},
\bauthor{\bsnm{{D{\'e}moulin}}, \binits{P.}},
\bauthor{\bsnm{{Mandrini}}, \binits{C.H.}},
\bauthor{\bsnm{{van Driel-Gesztelyi}}, \binits{L.}}:
\byear{2011},
\batitle{{Twisted Flux Tube Emergence Evidenced in Longitudinal Magnetograms:
  Magnetic Tongues}}.
\bjtitle{\solphys}
\bvolume{270},
\bfpage{45}.
\doiurl{https://doi.org/10.1007/s11207-011-9731-8}.
\adsurl{2011SoPh..270...45L}.
\end{barticle}
\endbibitem

\bibitem[\protect\citeauthoryear{{Manchester} et~al.}{2004}]{manchester2004}
\begin{barticle}
\bauthor{\bsnm{{Manchester}}, \binits{I.} \bsuffix{W.}},
\bauthor{\bsnm{{Gombosi}}, \binits{T.}},
\bauthor{\bsnm{{DeZeeuw}}, \binits{D.}},
\bauthor{\bsnm{{Fan}}, \binits{Y.}}:
\byear{2004},
\batitle{{Eruption of a Buoyantly Emerging Magnetic Flux Rope}}.
\bjtitle{\apj}
\bvolume{610},
\bfpage{588}.
\doiurl{https://doi.org/10.1086/421516}.
\adsurl{2004ApJ...610..588M}.
\end{barticle}
\endbibitem

\bibitem[\protect\citeauthoryear{{Mart{\'\i}nez-Sykora}, {Hansteen}, and
  {Carlsson}}{2008}]{MartinezSykora:2008}
\begin{barticle}
\bauthor{\bsnm{{Mart{\'\i}nez-Sykora}}, \binits{J.}},
\bauthor{\bsnm{{Hansteen}}, \binits{V.}},
\bauthor{\bsnm{{Carlsson}}, \binits{M.}}:
\byear{2008},
\batitle{{Twisted Flux Tube Emergence From the Convection Zone to the Corona}}.
\bjtitle{\apj}
\bvolume{679},
\bfpage{871}.
\doiurl{https://doi.org/10.1086/587028}.
\adsurl{2008ApJ...679..871M}.
\end{barticle}
\endbibitem

\bibitem[\protect\citeauthoryear{{Melrose}}{1995}]{melrose1995}
\begin{barticle}
\bauthor{\bsnm{{Melrose}}, \binits{D.B.}}:
\byear{1995},
\batitle{{Current Paths in the Corona and Energy Release in Solar Flares}}.
\bjtitle{\apj}
\bvolume{451},
\bfpage{391}.
\doiurl{https://doi.org/10.1086/176228}.
\adsurl{1995ApJ...451..391M}.
\end{barticle}
\endbibitem

\bibitem[\protect\citeauthoryear{{Parker}}{1996}]{parker1996a}
\begin{barticle}
\bauthor{\bsnm{{Parker}}, \binits{E.N.}}:
\byear{1996},
\batitle{{Comment on ``Current Paths in the Corona and Energy Release in Solar
  Flares''}}.
\bjtitle{\apj}
\bvolume{471},
\bfpage{489}.
\doiurl{https://doi.org/10.1086/177984}.
\adsurl{1996ApJ...471..489P}.
\end{barticle}
\endbibitem

\bibitem[\protect\citeauthoryear{{Poisson} et~al.}{2016}]{poisson2016}
\begin{barticle}
\bauthor{\bsnm{{Poisson}}, \binits{M.}},
\bauthor{\bsnm{{D{\'e}moulin}}, \binits{P.}},
\bauthor{\bsnm{{L{\'o}pez Fuentes}}, \binits{M.}},
\bauthor{\bsnm{{Mandrini}}, \binits{C.H.}}:
\byear{2016},
\batitle{{Properties of Magnetic Tongues over a Solar Cycle}}.
\bjtitle{\solphys}
\bvolume{291},
\bfpage{1625}.
\doiurl{https://doi.org/10.1007/s11207-016-0926-x}.
\adsurl{2016SoPh..291.1625P}.
\end{barticle}
\endbibitem

\bibitem[\protect\citeauthoryear{{Ravindra} et~al.}{2011}]{ravindra2011}
\begin{barticle}
\bauthor{\bsnm{{Ravindra}}, \binits{B.}},
\bauthor{\bsnm{{Venkatakrishnan}}, \binits{P.}},
\bauthor{\bsnm{{Tiwari}}, \binits{S.K.}},
\bauthor{\bsnm{{Bhattacharyya}}, \binits{R.}}:
\byear{2011},
\batitle{{Evolution of Currents of Opposite Signs in the Flare-productive Solar
  Active Region NOAA 10930}}.
\bjtitle{\apj}
\bvolume{740},
\bfpage{19}.
\doiurl{https://doi.org/10.1088/0004-637X/740/1/19}.
\adsurl{2011ApJ...740...19R}.
\end{barticle}
\endbibitem

\bibitem[\protect\citeauthoryear{{Sch\"ussler}}{1979}]{schuessler1979}
\begin{barticle}
\bauthor{\bsnm{{Sch\"ussler}}, \binits{M.}}:
\byear{1979},
\batitle{{Magnetic buoyancy revisited: analytical and numerical results for
  rising flux tubes.}}
\bjtitle{\aap}
\bvolume{71},
\bfpage{79}.
\adsurl{1979A&A....71...79S}.
\end{barticle}
\endbibitem

\bibitem[\protect\citeauthoryear{{Spruit}, {Title}, and {van
  Ballegooijen}}{1987}]{spruit1987}
\begin{barticle}
\bauthor{\bsnm{{Spruit}}, \binits{H.C.}},
\bauthor{\bsnm{{Title}}, \binits{A.M.}},
\bauthor{\bsnm{{van Ballegooijen}}, \binits{A.A.}}:
\byear{1987},
\batitle{{Is there a weak mixed polarity background field? Theoretical
  arguments}}.
\bjtitle{\solphys}
\bvolume{110},
\bfpage{115}.
\doiurl{https://doi.org/10.1007/BF00148207}.
\adsurl{1987SoPh..110..115S}.
\end{barticle}
\endbibitem

\bibitem[\protect\citeauthoryear{{Stein} and
  {Nordlund}}{2012}]{SteinNordlund:2012}
\begin{barticle}
\bauthor{\bsnm{{Stein}}, \binits{R.F.}},
\bauthor{\bsnm{{Nordlund}}, \binits{{\r{A}}.}}:
\byear{2012},
\batitle{{On the Formation of Active Regions}}.
\bjtitle{\apjl}
\bvolume{753},
\bfpage{L13}.
\doiurl{https://doi.org/10.1088/2041-8205/753/1/L13}.
\adsurl{2012ApJ...753L..13S}.
\end{barticle}
\endbibitem

\bibitem[\protect\citeauthoryear{{Syntelis} et~al.}{2019}]{Syntelis:2019}
\begin{barticle}
\bauthor{\bsnm{{Syntelis}}, \binits{P.}},
\bauthor{\bsnm{{Lee}}, \binits{E.J.}},
\bauthor{\bsnm{{Fairbairn}}, \binits{C.W.}},
\bauthor{\bsnm{{Archontis}}, \binits{V.}},
\bauthor{\bsnm{{Hood}}, \binits{A.W.}}:
\byear{2019},
\batitle{{Eruptions and flaring activity in emerging quadrupolar regions}}.
\bjtitle{\aap}
\bvolume{630},
\bfpage{A134}.
\doiurl{https://doi.org/10.1051/0004-6361/201936246}.
\adsurl{2019A&A...630A.134S}.
\end{barticle}
\endbibitem

\bibitem[\protect\citeauthoryear{{Takasao} et~al.}{2015}]{Takasao:2015}
\begin{barticle}
\bauthor{\bsnm{{Takasao}}, \binits{S.}},
\bauthor{\bsnm{{Fan}}, \binits{Y.}},
\bauthor{\bsnm{{Cheung}}, \binits{M.C.M.}},
\bauthor{\bsnm{{Shibata}}, \binits{K.}}:
\byear{2015},
\batitle{{Numerical Study on the Emergence of Kinked Flux Tube for
  Understanding of Possible Origin of {\ensuremath{\delta}}-spot Regions}}.
\bjtitle{\apj}
\bvolume{813},
\bfpage{112}.
\doiurl{https://doi.org/10.1088/0004-637X/813/2/112}.
\adsurl{2015ApJ...813..112T}.
\end{barticle}
\endbibitem

\bibitem[\protect\citeauthoryear{{Titov} and {D{\'e}moulin}}{1999}]{titov1999}
\begin{barticle}
\bauthor{\bsnm{{Titov}}, \binits{V.S.}},
\bauthor{\bsnm{{D{\'e}moulin}}, \binits{P.}}:
\byear{1999},
\batitle{{Basic topology of twisted magnetic configurations in solar flares}}.
\bjtitle{\aap}
\bvolume{351},
\bfpage{707}.
\adsurl{1999A&A...351..707T}.
\end{barticle}
\endbibitem

\bibitem[\protect\citeauthoryear{{Titov}, {Priest}, and
  {Demoulin}}{1993}]{titov1993}
\begin{barticle}
\bauthor{\bsnm{{Titov}}, \binits{V.S.}},
\bauthor{\bsnm{{Priest}}, \binits{E.R.}},
\bauthor{\bsnm{{Demoulin}}, \binits{P.}}:
\byear{1993},
\batitle{{Conditions for the appearance of ``bald patches'' at the solar
  surface}}.
\bjtitle{\aap}
\bvolume{276},
\bfpage{564}.
\adsurl{1993A&A...276..564T}.
\end{barticle}
\endbibitem

\bibitem[\protect\citeauthoryear{{Toriumi} and {Hotta}}{2019}]{Toriumi:2019}
\begin{barticle}
\bauthor{\bsnm{{Toriumi}}, \binits{S.}},
\bauthor{\bsnm{{Hotta}}, \binits{H.}}:
\byear{2019},
\batitle{{Spontaneous Generation of {\ensuremath{\delta}}-sunspots in
  Convective Magnetohydrodynamic Simulation of Magnetic Flux Emergence}}.
\bjtitle{\apjl}
\bvolume{886},
\bfpage{L21}.
\doiurl{https://doi.org/10.3847/2041-8213/ab55e7}.
\adsurl{2019ApJ...886L..21T}.
\end{barticle}
\endbibitem

\bibitem[\protect\citeauthoryear{{Toriumi} et~al.}{2020}]{Toriumi:2020}
\begin{barticle}
\bauthor{\bsnm{{Toriumi}}, \binits{S.}},
\bauthor{\bsnm{{Takasao}}, \binits{S.}},
\bauthor{\bsnm{{Cheung}}, \binits{M.C.M.}},
\bauthor{\bsnm{{Jiang}}, \binits{C.}},
\bauthor{\bsnm{{Guo}}, \binits{Y.}},
\bauthor{\bsnm{{Hayashi}}, \binits{K.}},
\bauthor{\bsnm{{Inoue}}, \binits{S.}}:
\byear{2020},
\batitle{{Comparative Study of Data-driven Solar Coronal Field Models Using a
  Flux Emergence Simulation as a Ground-truth Data Set}}.
\bjtitle{\apj}
\bvolume{890},
\bfpage{103}.
\doiurl{https://doi.org/10.3847/1538-4357/ab6b1f}.
\adsurl{2020ApJ...890..103T}.
\end{barticle}
\endbibitem

\bibitem[\protect\citeauthoryear{{T{\"o}r{\"o}k} and {Kliem}}{2003}]{torok2003}
\begin{barticle}
\bauthor{\bsnm{{T{\"o}r{\"o}k}}, \binits{T.}},
\bauthor{\bsnm{{Kliem}}, \binits{B.}}:
\byear{2003},
\batitle{{The evolution of twisting coronal magnetic flux tubes}}.
\bjtitle{\aap}
\bvolume{406},
\bfpage{1043}.
\doiurl{https://doi.org/10.1051/0004-6361:20030692}.
\adsurl{2003A&A...406.1043T}.
\end{barticle}
\endbibitem

\bibitem[\protect\citeauthoryear{{T{\"o}r{\"o}k} et~al.}{2014}]{torok2014}
\begin{barticle}
\bauthor{\bsnm{{T{\"o}r{\"o}k}}, \binits{T.}},
\bauthor{\bsnm{{Leake}}, \binits{J.E.}},
\bauthor{\bsnm{{Titov}}, \binits{V.S.}},
\bauthor{\bsnm{{Archontis}}, \binits{V.}},
\bauthor{\bsnm{{Miki{\'c}}}, \binits{Z.}},
\bauthor{\bsnm{{Linton}}, \binits{M.G.}},
\bauthor{\bsnm{{Dalmasse}}, \binits{K.}},
\bauthor{\bsnm{{Aulanier}}, \binits{G.}},
\bauthor{\bsnm{{Kliem}}, \binits{B.}}:
\byear{2014},
\batitle{{Distribution of Electric Currents in Solar Active Regions}}.
\bjtitle{\apjl}
\bvolume{782},
\bfpage{L10}.
\doiurl{https://doi.org/10.1088/2041-8205/782/1/L10}.
\adsurl{2014ApJ...782L..10T}.
\end{barticle}
\endbibitem

\bibitem[\protect\citeauthoryear{{Vemareddy}}{2019}]{vemareddy2019}
\begin{barticle}
\bauthor{\bsnm{{Vemareddy}}, \binits{P.}}:
\byear{2019},
\batitle{{Degree of electric current neutralization and the activity in solar
  active regions}}.
\bjtitle{\mnras}
\bvolume{486},
\bfpage{4936}.
\doiurl{https://doi.org/10.1093/mnras/stz1020}.
\adsurl{2019MNRAS.486.4936V}.
\end{barticle}
\endbibitem

\bibitem[\protect\citeauthoryear{{Venkatakrishnan} and
  {Tiwari}}{2009}]{venkatakrishnan2009}
\begin{barticle}
\bauthor{\bsnm{{Venkatakrishnan}}, \binits{P.}},
\bauthor{\bsnm{{Tiwari}}, \binits{S.K.}}:
\byear{2009},
\batitle{{On the Absence of Photospheric Net Currents in Vector Magnetograms of
  Sunspots Obtained from Hinode (Solar Optical
  Telescope/Spectro-Polarimeter)}}.
\bjtitle{\apjl}
\bvolume{706},
\bfpage{L114}.
\doiurl{https://doi.org/10.1088/0004-637X/706/1/L114}.
\adsurl{2009ApJ...706L.114V}.
\end{barticle}
\endbibitem

\bibitem[\protect\citeauthoryear{{Wheatland}}{2000}]{wheatland2000}
\begin{barticle}
\bauthor{\bsnm{{Wheatland}}, \binits{M.S.}}:
\byear{2000},
\batitle{{Are Electric Currents in Solar Active Regions Neutralized?}}
\bjtitle{\apj}
\bvolume{532},
\bfpage{616}.
\doiurl{https://doi.org/10.1086/308577}.
\adsurl{2000ApJ...532..616W}.
\end{barticle}
\endbibitem

\bibitem[\protect\citeauthoryear{{Yelles Chaouche} et~al.}{2009}]{Yelles:2009}
\begin{barticle}
\bauthor{\bsnm{{Yelles Chaouche}}, \binits{L.}},
\bauthor{\bsnm{{Cheung}}, \binits{M.C.M.}},
\bauthor{\bsnm{{Solanki}}, \binits{S.K.}},
\bauthor{\bsnm{{Sch{\"u}ssler}}, \binits{M.}},
\bauthor{\bsnm{{Lagg}}, \binits{A.}}:
\byear{2009},
\batitle{{Simulation of a flux emergence event and comparison with observations
  by Hinode}}.
\bjtitle{\aap}
\bvolume{507},
\bfpage{L53}.
\doiurl{https://doi.org/10.1051/0004-6361/200913181}.
\adsurl{2009A&A...507L..53Y}.
\end{barticle}
\endbibitem

\end{thebibliography}


\clearpage



\end{article} 

\end{document}